\documentclass[aps,pre,twocolumn,superscriptaddress,showpacs,showkeys ]{revtex4-2}

\usepackage{graphicx}                   
\usepackage{hyperref}                   
\usepackage{amsmath,amssymb}            
\usepackage{epstopdf}
\usepackage{subcaption}					

\usepackage{color}
\definecolor{orange}{rgb}{1,0.5,0}
\definecolor{brown}{rgb}{0.65, 0.16, 0.16}
\definecolor{phlox}{rgb}{0.87, 0.0, 1.0}

\begin{document}
	\title{Statistical analysis of the drying pattern of coffee}
	\author{J. Cheraghalizadeh}
	\affiliation{Department of Physics, University of Mohaghegh Ardabili, P.O. Box 179, Ardabil, Iran}
	\author{S. Tizdast}
	\affiliation{Department of Physics, University of Mohaghegh Ardabili, P.O. Box 179, Ardabil, Iran}
	\author{N. Valizadeh}
	\affiliation{Department of Physics, K.N. Toosi University of Technology, Tehran 15875-4416, Iran}
	\author{S. Doostdari}
	\affiliation{Department of Physics, University of Mohaghegh Ardabili, P.O. Box 179, Ardabil, Iran}
	\author{M. N. Najafi}
	\affiliation{Department of Physics, University of Mohaghegh Ardabili, P.O. Box 179, Ardabil, Iran}
	\email{morteza.nattagh@gmail.com}
	
	\begin{abstract}	
The influence of sugar on the geometrical properties of dried coffee stain patterns is experimentally investigated. The sugar content is quantified by its relative mass, $m$. To analyze this effect, the dried stain, formed due to Marangoni flow, is mapped onto a rough surface, and various statistical tools are employed to study crack statistics. These analyses encompass both local quantities, such as mass distribution, and global quantities, including the gyration radius and loop length of the dried clusters. The global quantities pertain to the coffee ring, defined as the boundary of the dried coffee stain. For sufficiently large values of $m$ the observed exponents approach those of the Gaussian free field (GFF). In this regime, the loop fractal dimension is approximately $\frac{3}{2}$, while the loop and gyration radius distribution exponents align with $\tau_l=\frac{7}{3}$ and $\tau_r=3$, respectively. Multifractal analysis of the mass configuration in the dried pattern reveals that the mass-fractal dimension is $1.76\pm 0.04$ for the $m=0$, decreasing as $m$ increases. This trend can be attributed to the droplet becoming more hydrophilic, leading to sparser spatial patterns, a conclusion supported by contact angle analysis.
\end{abstract}
\pacs{05., 05.20.-y, 05.10.Ln, 05.45.Df}
\keywords{Drying pattern, separating lines, statistical analysis, complex fluids}
\maketitle
	
	\section{Introduction}
Complex fluids are a category of fluids that display unconventional mechanical behavior under applied stress or strain, driven by the geometric constraints associated with phase coexistence. Non-Newtonian fluids are a key example, characterized by a non-linear relationship between shear stress and flow rate. These fluids are generally mixtures containing two phases—such as solid-liquid, liquid-liquid, solid-gas, or liquid-gas—often involving suspensions or solutions of macromolecules~\cite{saramito2016complex}. The dynamics of complex fluids, particularly with regards to drying, have long posed a challenge in the field of fluid dynamics~\cite{sefiane2014patterns}.  When a liquid evaporates around a droplet, an outward fluid flow is required to maintain a wet surface, resulting in particle accumulation in the contact line. According to \cite{deegan1997capillary}, the pinning in the contact line is caused by the accumulation of solid components in that area, known as self-pinning. The dynamics and drying patterns of complex fluids are strongly influenced by the presence of pinning centers. Contact line pinning, combined with evaporation, provides the necessary conditions for ring formation~\cite{deegan2000pattern}. Marangoni flow, a convective flow driven by surface tension gradients within the droplet, plays a critical role in shaping the drying pattern alongside evaporation and wettability~\cite{zeid2013influence,chen2016blood,chen2016wetting,sefiane2014patterns,hertaeg2021pattern,shen2010minimal}. The non-uniform evaporation process creates a temperature gradient and, consequently, a surface tension gradient that generates a Marangoni flow. The flow direction is determined by the relative conductivities of the substrate and liquid \cite{ristenpart2007influence}. These patterns have been observed and analyzed in nanofluids \cite{brutin2013influence} and polymers \cite{sefiane2014patterns} and blood~\cite{brutin2013influence}, the latter being an important example of complex fluid with significant Marangoni effect~\cite{chen2016blood,bahmani2017study,muniandy2013non,brutin2011pattern,zeid2013influence2,sefiane2014patterns,chen2016blood,chen2016wetting,pal2020concentration}. The statistical analysis of crack patterns in dried complex fluids is particularly significant to recover their hidden aspects, e.g. for blood it aids in diagnosing blood-related conditions like anemia~\cite{brutin2011pattern} and thalassemia~\cite{bahmani2017study}. While Marangoni flow is thought to play a crucial role in the formation of cracks within dried patterns, the precise nature of this relationship remains poorly understood~\cite{mondal2023physics}. \\ 

Coffee is a notable example of a complex fluid, sharing similarities with blood~\cite{brutin2011pattern}. Numerous studies have investigated how the evaporation of thin films and small droplets can be utilized for particle self-assembly and surface patterning, drawing parallels to the coffee ring effect. Deegan \textit{et al.} \cite{deegan1997capillary,deegan2000pattern,deegan2000contact} made a significant contribution to understanding the self-pinning phenomena of coffee stain rings and the resulting dried pattern of complex droplets. They attributed the formation of ring-like stains to an outward flow that occurs during the drying dynamics of the droplet. The outward capillary flow causes the ring mass to increase following a power-law, which is anticipated to impact processes such as printing, washing, and coating \cite{tekin2008inkjet,derby2010inkjet}. Along with the roughness or heterogeneity of the underlying surface, which affects the pinning properties, the Marangoni effect is also critical to the formation properties and the pinned ring in coffee. Additionally, it has been demonstrated that an inward flow towards the center of the droplet can also be induced, depending on the evaporative driving force \cite{fischer2002particle}.  The evaporative mass flux is determined by the contact angle and is predicted to diverge at the contact line. In \cite{fischer2002particle}, the problem was addressed by applying a lubrication approximation to the Navier-Stokes equations, and several evaporative flux modes were examined. An analytical model for the outward flux and contact line formation was proposed in \cite{popov2005evaporative}, which is based on the coexistence of liquid and deposit phases. The model's outcomes were found to be universal, meaning they are not dependent on any free or fitting parameters. Solvent evaporation from a capillary bridge produces a novel pattern of concentric rings with a gradient, as reported in \cite{xu2006self}. It has been shown that the formation of coffee rings depends not only on a pinned contact line but also on the suppression of Marangoni flow. Marangoni flow can reverse the deposition process by moving particles from the edge to the center, leading to the formation of a coffee ring at the center~\cite{hu2006marangoni}. The elimination of coffee-rings which has a lot of applications in coatings in printing~\cite{park2006control,tekin2008inkjet,derby2010inkjet}, and biology~\cite{dugas2005droplet}, is possible via tuning the shape of the suspended particles~\cite{yunker2011suppression}, as well as the circulating radial Marangoni flows (Marangoni eddy)~\cite{still2012surfactant}. A minimal size of the coffee-rings were considered in~\cite{shen2010minimal}.\\

Various drying patterns have been found as a result of the interaction of colloidal particles with substrate, i.e. when the solute particles in a solution remain attached to the solid surface due to colloidal droplet evaporation. Several systems have been found to exhibit various behaviors, including CoPt$_3$ particles with a bilayer structure~\cite{govor2004nanoparticle}, and liquid crystal pattern formation in drying droplets of DNA~\cite{smalyukh2006structure}. Many studies have been conducted to examine the effect of the experimental conditions of the patters, like the effect of substrate conductivity~\cite{ristenpart2007influence}, the reverse coffee-ring effect by laser-induced differential evaporation~\cite{yen2018reversing}, spiral collide deposition~\cite{zang2019evaporation}, and drying pattern of colloidal suspensions on the inclined substrates~\cite{thampi2020beyond}. This problem is fascinating from several viewpoints, especially considering its relevance to printing~\cite{tekin2008inkjet,derby2010inkjet}. For instance, the dynamics of the process have been investigated in studies such as~\cite{craster2009dynamics,bonn2009wetting,han2012learning,erbil2012evaporation,brutin2018recent,mampallil2018review,al2019new,zang2019evaporation}. Additionally, a phase diagram for the self-assembly of colloidal particles in the dried pattern was proposed in~\cite{bhardwaj2010self}. Although there is a vast body of literature on the dried pattern of coffee, the statistics of cracks within the bulk of the dried droplets have received little attention. In our study, we not only investigate the classical statistics of cracks, but also the density statistics of deposited collides in the dried coffee droplets using multifractal analysis. Sugar amount is an external parameter that affects the concentration of coffee. Our research demonstrates that this system exhibits robust scale-invariant properties. The exponents associated with global statistical measures, such as gyration radius and loop lengths, remain unchanged despite variations in the sugar concentration. However, the multifractal properties of the system are dependent on the amount of sugar present. Section~\ref{SEC:Experminet} is devoted to the description of our experimental set up and details. \\

The paper flow goes as follows: In the next section we review some aspects of the dried pattern of colloidal complex fluids. A basic theory of multifractal analysis is presented in SEC.~\ref{SEC:MA}. We describe the results of the multifractal analysis on the experimental outcomes in Sec.~\ref{SEC:Res}. We close the paper by a conclusion.
 
\section{The drying pattern of colloidal complex fluids}
Droplet drying is a complex dynamic process in which solute evaporation and Marangoni flow play dominant roles. Additional factors influencing this process include surface tension, substrate wettability, droplet contact angle, and hydrodynamic interactions. The distinctive behavior of coffee particles in water arises from their polar properties. Coffee molecules have hydrophilic heads that dissolve in water and hydrophobic tails that repel it. When introduced to water, the hydrophilic heads dissolve while the hydrophobic tails separate from the water, causing the molecules to accumulate at the water's surface and effectively reduce surface tension~\cite{bird1987dynamics}. The shape that droplets assume when placed on a surface is a physical manifestation of surface tension and varies according to the level of surface tension present. Droplets may either spread out and wet the surface or remain as a distinct droplets depending on the surface tension. This shape is determined by minimizing the surface energy at the three boundaries: the interface between the droplet's fluid and the surrounding air, between the fluid and the solid surface, and between the air and the solid surface. The formation of the contact angle (the angle between the droplet's tangential line and the substrate at the contact line) is a useful criterion for differentiating between different phases and is directly related to the surface's wettability.  On hydrophilic surfaces, the contact angle is smaller than $\frac{\pi}{2}$, while on hydrophobic surfaces, it is larger than $\frac{\pi}{2}$. In the asymptotic limit, the contact angle approaches $0$ for hydrophilic surfaces and $\pi$ for hydrophobic surfaces~\cite{bird1987dynamics}.\\
During the formation of a coffee ring, particle evaporation is a hydrodynamic process that disperses solid particles along a horizontal line. Once the liquid evaporates, the precipitated ring sediment remains on the substrate, containing all the solvents. The droplets create the necessary conditions for the formation of the ring, which begins with the pinning of the contact line. The pinned line is a fixed-line located at the boundary of the droplet, separating the dry and wet regions, beyond which particles cannot move. The accumulation of solid particles in the contact line, along with any irregularities they create, results in pinning, preventing the contact line from moving and causing a ring to form at the pinned line, and evaporation begins from this ring. The Marangoni effect, which is linked to the surface tension gradient in the interface between two fluids, is a phenomenon that directly concerns the minimization of surface energy. In our study, the Marangoni effect plays a crucial role in the movement of colloidal particles and involves mass transfer due to the surface tension gradient at the boundary between two fluids. The velocity of the particles can be expressed as follows:~\cite{roche2014marangoni,gaston2021marangoni,keiser2017marangoni}
	\begin{equation}
		u=\frac{(\sigma_1-\sigma_2)^{\frac{2}{3}}}{(\zeta\rho)^{1/3}r^{1/3}},
	\end{equation}
where, $\sigma_{1,2}$ represents the surface tension of fluids 1 and 2 (assuming $\sigma_1>\sigma_2$), while $\rho$ and $\zeta$ represent the mass density and viscosity of fluid 1. The diameter of the growing stain is denoted by $r$. During the evaporation of the liquid and the formation of a coffee ring, the Marangoni effect occurs and causes colloidal particles to be transported to the outer edge of the droplet. The phenomenon of self-pinning, also known as contact line pinning, was suggested as a primary factor in the hydrodynamic mechanism that leads to the formation of rings. This occurs as the collides are transported toward the contact line ~\cite{deegan2000pattern}. The reason why a droplet on a vertical surface can resist the force of gravity is due to the contact line being anchored to the irregularities present on the host substrate. There is an anticipated correlation between the statistics of the contact line and self-pinning and the contact angle. This is because both of these phenomena are associated with the propensity of colloidal particles to stick to the substrate. In this paper, we examine another crucial question concerning the statistics of particles deposited within the bulk of the droplet, which may contain randomly formed cracks. Similar cracks are observed in other dried complex droplets, such as blood~\cite{brutin2011pattern, bahmani2017study}, highlighting the importance of this phenomenon. This underscores the potential for classifying complex fluids based on the statistical analysis of these patterns. In the case of scaling behaviors, the critical exponents play a crucial role in achieving this objective by mapping the problem to a standard class in critical phenomena~\cite{biswas2018drying,thiele2014patterned}. The hydrodynamic interaction was also considered in literature, see~\cite{bird1987dynamics} for example. As demonstrated in the subsequent analysis, the scaling properties are observed in the case study presented in this paper.

\section{Multifractal Analysis (MA)}\label{SEC:MA}
This section introduces the multifractal analysis (MA) approach for systems containing partially filled (black) pixels.  In single fractal systems, the system is described using a single set of exponents, which is not possible for multifractal systems. Consider a two-dimensional image containing a self-similar pattern represented by black and white pixels. The MA approach involves partitioning the image's space into boxes, each with a size of $\delta$. The number of black pixels inside the $i$th box of size $\delta$, is denoted by $N_i(\delta)$, and is used to calculate the filling fraction of the box. This fraction is expressed as the ratio of the number of black pixels to the size of the box as follows:
	\begin{equation}
		{\mu _i}(\delta ) = \frac{{{N_i}(\delta )}}{N},
	\end{equation}
	where $N$ is the total number of pixels. MA analysis is based in defining a $q$-generalized partition function:
	\begin{equation}
		Z_q(\delta ) = \sum\limits_i {{{[{\mu _i}(\delta )]}^q}},
	\end{equation}
where the variable $q$ represents a moment. In scale-invariant systems, $Z_q$ exhibits power-law scaling with respect to $\delta$, although the exponent may not be constant across all scales. For unique, well-defined exponents in the thermodynamic limit, we are interested to small scales $\delta\rightarrow 0$. Specifically, we define the exponent $\tau_q$ as follows:
	\begin{equation}
		\tau_q = \mathop {\lim }\limits_{\delta  \to 0} \frac{{{{\log }_2}{Z_q}(\delta )}}{{{{\log }_2}\delta }}.
	\end{equation}
The $q-$generalized fractal dimension is then defined as
	\begin{equation}
		{D_q} = \frac{\tau_q}{{q - 1}},
		\label{eq73}
	\end{equation}
which defines an infinite series of fractal dimensions. An important limit is called mass fractal dimension is $q\rightarrow 0$, \begin{equation}
	D_f = \lim_{q\rightarrow 0} D_q,
	\label{Eq:FDMA}
\end{equation} 
and also the (Shannon) information dimension $q\rightarrow 1$ \cite{taniguchi2003regulatory}
	\begin{equation}
		{D_1} = \mathop {\lim }\limits_{\delta  \to 0} \frac{{\sum\limits_i {{\mu _i}(\delta ){{\log }_2}{\mu _i}(\delta )} }}{{{{\log }_2}\delta }}.
	\end{equation}
Note that the numerator of the above equation is exactly the information entropy of the system, while Eq.~\ref{eq73} is the $q$ Renyi entropy. For $q=2$, the numerator is called the correlation dimension. For more information on the definitions and interpretations of an infinite series of fractal dimensions, see Appendix~\ref{SEC:Appendix}.\\
	
For multifractal systems, the exponents are changing place to place or scale to scale. To quantify the spectrum of the exponents, one has to apply a new function which is Legendre transformation of $\tau_q$. This spectrum of the exponents $f$ is defined as follows
\begin{equation}
	f({\alpha _q}) = q{\alpha _q} - {\tau _q},
	\label{Eq:Legendre}
\end{equation}
where $\alpha_q = \frac{d\tau_q }{dq}$. In general, $f(\alpha)$ is a non-linear function of $\alpha$, the peak of which determines the average exponent $\bar{\tau}_q$, and the maximum and the minimum shows the range of exponents. For more details see SEC.~\ref{SEC:Multi}. \\

To exploit the  rough surface techniques, we utilized a mapping according to which the color of each segment of a photo is interpreted as a local ``height'', and based on certain threshold values extract the contour lines. This process transforms the photos into a collection of loops, collectively referred to as the contour loop ensemble (CLE), where the iso-lines corresponded to the intensity cut values determined by the thresholds. After generating the CLE, standard statistical techniques were applied to analyze the data. For each loop, its length and radius of gyration were measured and utilized to investigate the statistical characteristics, including the fractal dimension (for which the data exhibits scale-invariance). The radius of gyration is defined as follows:
\begin{equation}
	{r^2} = \frac{1}{l}\sum\limits_{i = 1}^l \left| {{\vec r}_i} - {{\vec r}_{\text{com}}}\right|^2,
\end{equation}
where $l$ the length of the loop and ${\vec r}_{\text{com}} $ the center of mass of the loop ${\vec r_{\text{com}}} = \frac{1}{l}\sum\limits_{i = 1}^l \vec r_i $. For scale-invariant systems, the scaling relation between $l$ and $r$ gives the fractal dimension $\gamma_{lr}$, defined by
\begin{equation}
	\left\langle {\log (l)} \right\rangle  = \gamma_{lr}\left\langle {\log (r)} \right\rangle + \text{constant},
	\label{Eq:FD}
\end{equation}
where $\gamma_{lr}$ is the fractal dimension of the loops, and $\left\langle \right\rangle $ means the ensemble average. Also, for scale-invariant systems, one expects that the distribution functions show power-law behaviors
\begin{equation}
	P_x(x)\propto x^{-\tau_x}
	\label{Eq:distribution}
\end{equation}
where $x=l,r$.  It is important to note that this equation is valid only within a certain spatial scale, above which finite-size effects become significant.
	
\section{Experiments on the Dried Coffee pattern}~\label{SEC:Experminet}

In this section we describe the experimental set up and details (in the following subsection). In the next subsection we present the results.
\subsection{Experimental details}
To minimize the impact of irregularities in the background and their effect on the statistics of the dried patterns, we used mica sheets as a smooth and homogeneous substrate for the coffee droplets in this study. It's worth noting that disorder in the substrate can cause pinning, as previously observed~\cite{deegan1997capillary}. The mica sheets were cleaned with distilled water and allowed to air-dry, a process that typically takes less than 20 minutes at room temperature. The coffee suspension consisted of $1\pm 10^{-6}$ grams of Turkish coffee in $30\pm 1$ grams of water, with particle size set at $0.1\pm 0.01$ micrometers. We used a pipette with a precision of $0.1$ mL to deposit coffee droplets onto the mica substrate. To capture images, we utilized a Canon digital camera with a full HD resolution and a $1$ to $5$ macro lens. We employed an adjustable stand holder to position the camera at a variable distance from the samples. All samples were prepared in nearly identical environmental conditions. To ensure uniform background illumination, photos were taken at a fixed time of day, and a diffused fluorescent light source was used in the background. The preparation of the samples was conducted under ambient conditions with a temperature of $23 \pm 1$ degrees Celsius. To ensure proper evaporation of the solvent, it was left in the open air for one hour before the application of the droplets onto the mica sheets. Each sample was generated by adding $0.5$ ml of solvent onto the substrate, resulting in a circular stain with an estimated radius of approximately $2 cm$. To investigate the impact of sugar, we introduced different quantities of sugar, specifically $m=1.5$, $2$, and $2.5$ grams, into a coffee water solution with a ratio of $1:30$ grams. The subsequent steps remained unchanged from the sugar-free experiment. Sugar has been observed to alter the ratio of adhesion energies, resulting in an increase in adhesion. This effect is demonstrated by the darker appearance of dried stains in samples that contain sugar, as compared to those without sugar. In total, we generated $60$ samples for each condition. To examine the overall characteristics of the dried pattern, we extracted image segments measuring $2000 \times 2000$ pixels from the center of the coffee stain. These images, which consist of rescaled red, blue, and green matrices, were then converted into a grayscale matrix. In order to establish a standardized metric, we adjusted the photo intensities by rescaling them and dividing them by their average intensity. 
	\begin{figure*}
		\begin{subfigure}{0.5\textwidth}\includegraphics[width=\textwidth]{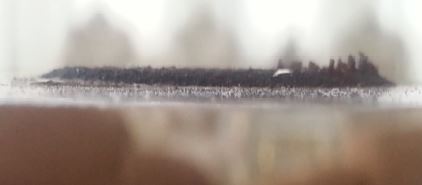}
			\caption{}
			\label{C33}
		\end{subfigure}
		\begin{subfigure}{0.32\textwidth}\includegraphics[width=\textwidth]{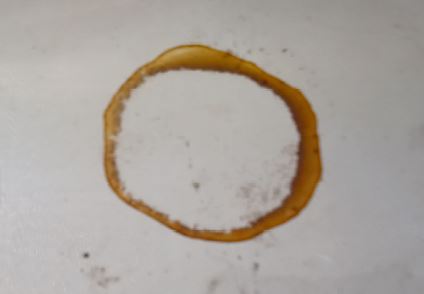}
			\caption{}
			\label{C23}
		\end{subfigure}
		\caption{(a) Concentration of colloidal particles in the center of the loop. (b) Coffee ring formed by the deposited particles on the edge of a coffee spot.}
	\end{figure*} 
	
The outward motion of particles, as predicted in the theory causes the formation of coffee-ring as depicted in Fig.~\ref{C33} (side view of a completely dried droplet in which the particles have adhered to the substrate) and~\ref{C23} (top view). Typically, the drying process of colloidal coffee suspensions can be divided into three temporal stages. During the first $20$ minutes of the experiment, the particles are advected to the contact line where they accumulate around the ring, while the liquid gently evaporates. Once the contact line of the ring has been established, the droplet's radius will vary depending on the amount of solvent used. In the second stage, which lasts approximately one hour, the volume of coffee inside the ring area decreases, leaving behind a thin layer of solvent. This process causes the stain's color to lighten. Additionally, during this stage, the contact line advances to the point of complete pinning, as depicted in Fig~\ref{C23}. In the third and final stage, which lasts for approximately $30$ minutes, the stain is fully dried, and the inner part of the ring begins to take shape. This process is accompanied by the formation of cracks, and the color of the central portion becomes lighter than that of the halo's edge. Our statistical measurements are conducted during this stage. We explore the impact of sugar on the drying process and discover that the crack statistics are influenced by the quantity of sugar used.

\section{Results}~\label{SEC:Res}
\begin{figure}
	\centering
	\includegraphics[scale=0.4]{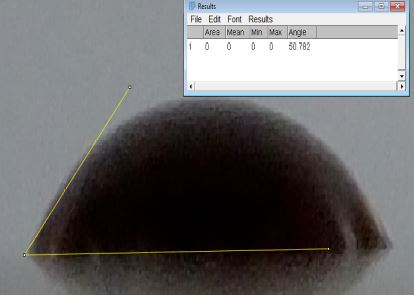}
	\caption{Side view of coffee drops, and the corresponding contact angle.}
	\label{Fig:C83}
\end{figure}
\begin{figure*}
		\begin{subfigure}{0.30\textwidth}\includegraphics[width=\textwidth]{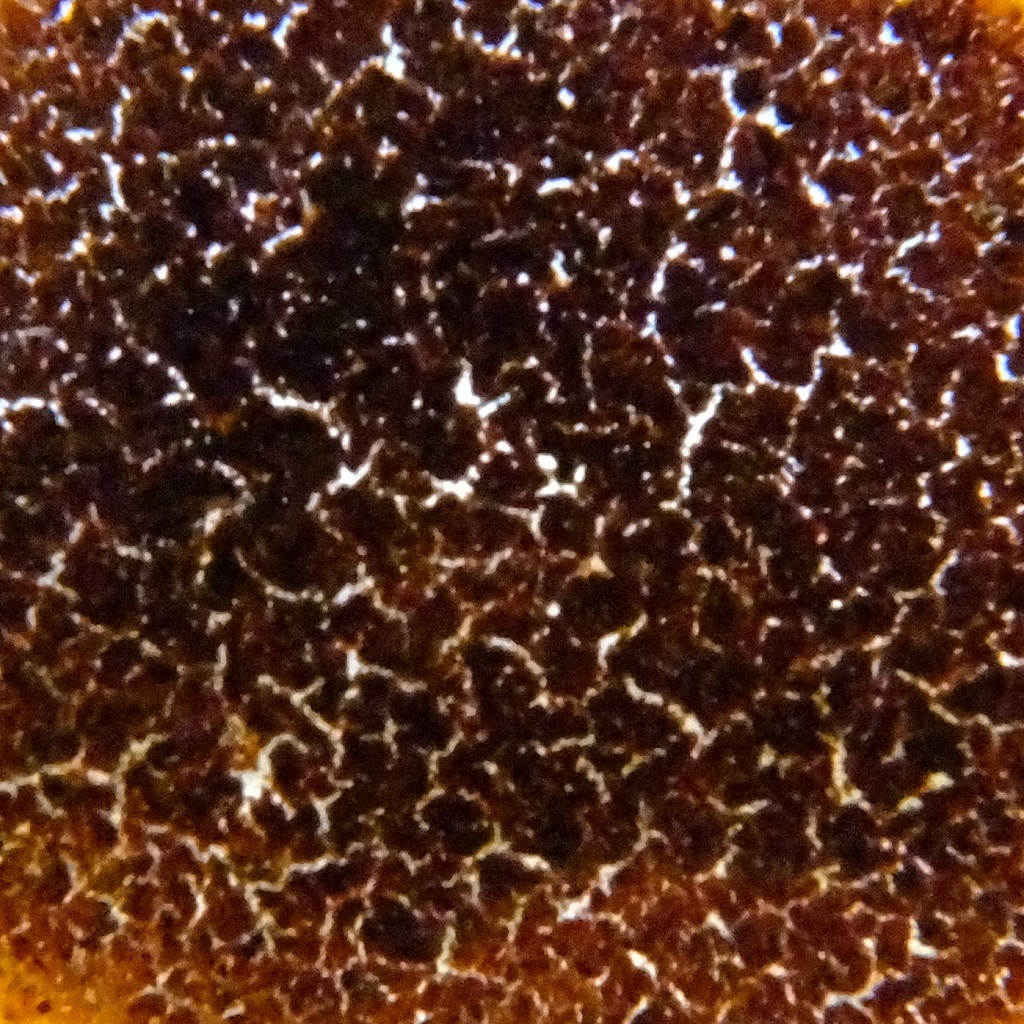}
			\caption{}
			\label{Fig:C93}
		\end{subfigure}
		\begin{subfigure}{0.40\textwidth}\includegraphics[width=\textwidth]{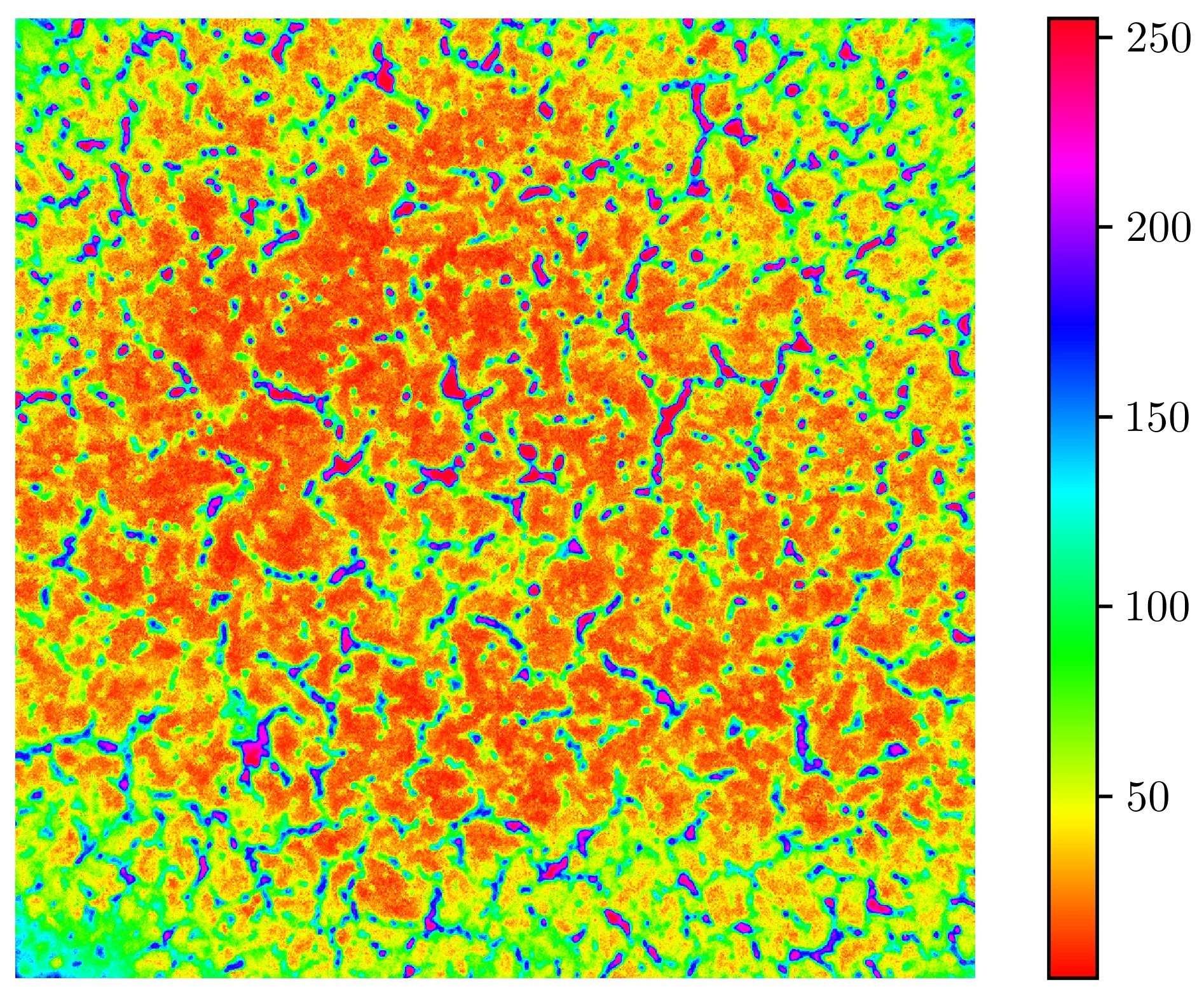}
			\caption{}
			\label{Fig:capture}
		\end{subfigure}
		\begin{subfigure}{0.73\textwidth}\includegraphics[width=\textwidth]{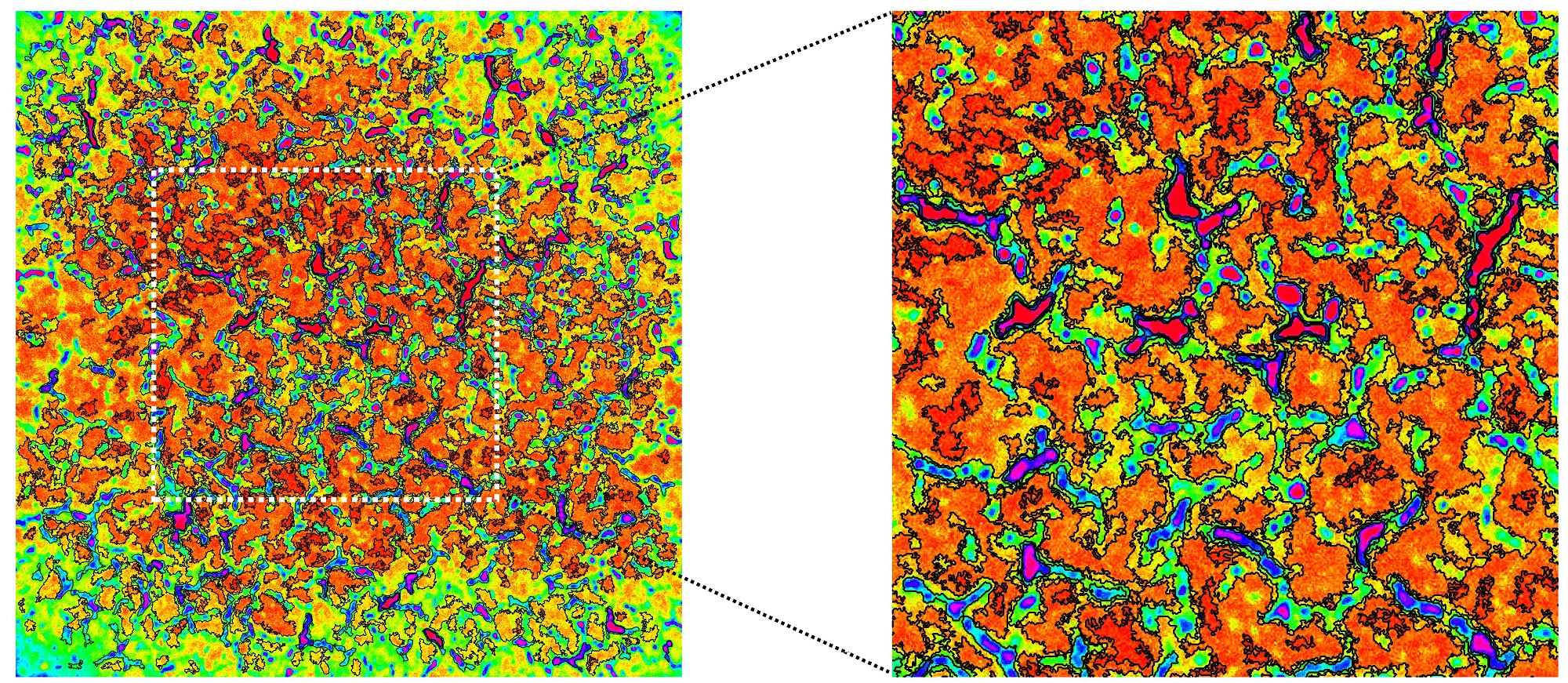}
			\caption{}
			\label{Fig:C113}
		\end{subfigure}
		\caption{The cropped part of the image of a dried coffee pattern.  Form a loop of squares of the same intensity. (b)Image of created loops}
	\end{figure*}

The contact angle of a small droplet has been shown in Fig.~\ref{Fig:C83}, in which the contact angle was obtained using the ImageJ software~\footnote{ImageJ software bundled with 64-bit Java 8}. Our observations indicate that the contact angle of a sugar-free coffee droplet is $50 \pm 3$ degrees, whereas for coffee with sugar, it is $48\pm3$ degrees (with negligible dependence on the amount of sugar within the limits of our experimental accuracy). These findings indicate that the addition of sugar causes a reduction in the contact angle, which is consistent with the fact that sugar makes the coffee hydrophilic. A sample with cracks is shown in Fig.~\ref{Fig:C93}, the intensity field of which is shown in Fig.~\ref{Fig:capture}. The CLE is obtained by cutting such figures from specific thresholds. The method of preparing the figures (cropping from the central parts) is shown in Fig.~\ref{Fig:C113}.    

\begin{figure*}
	\begin{subfigure}{0.47\textwidth}\includegraphics[width=\textwidth]{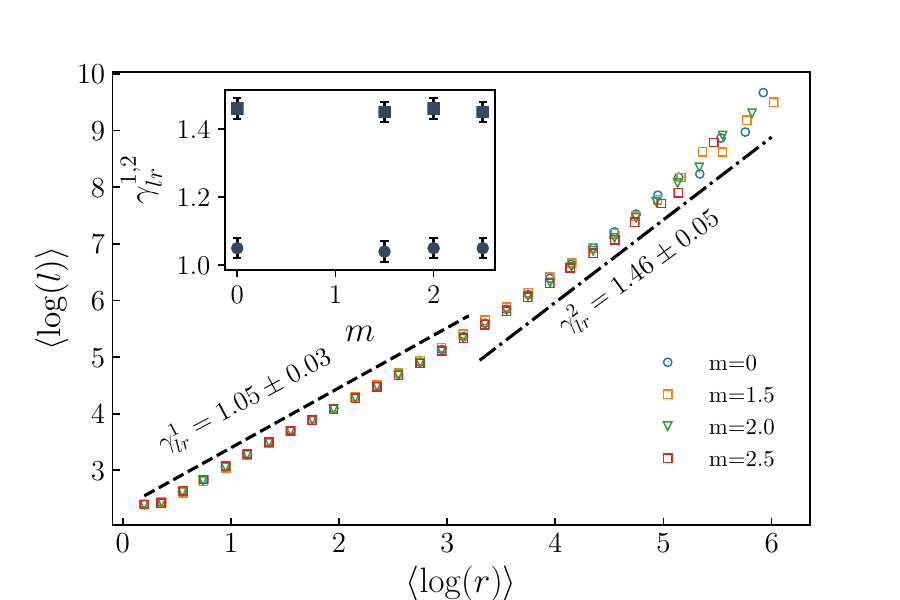}
		\caption{}
		\label{C123}
	\end{subfigure}
	\begin{subfigure}{0.46\textwidth}\includegraphics[width=\textwidth]{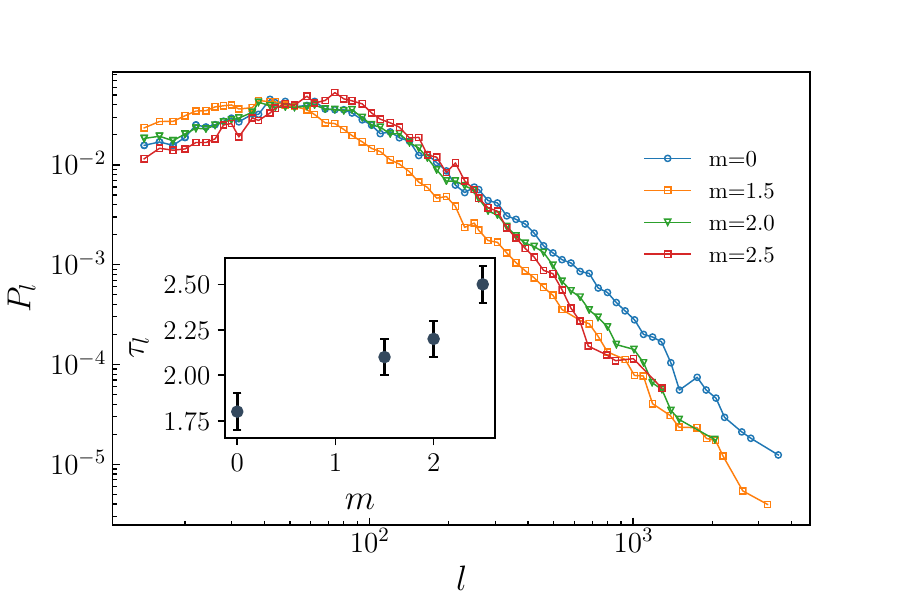}
		\caption{}
		\label{C133}
	\end{subfigure}
	\begin{subfigure}{0.47\textwidth}\includegraphics[width=\textwidth]{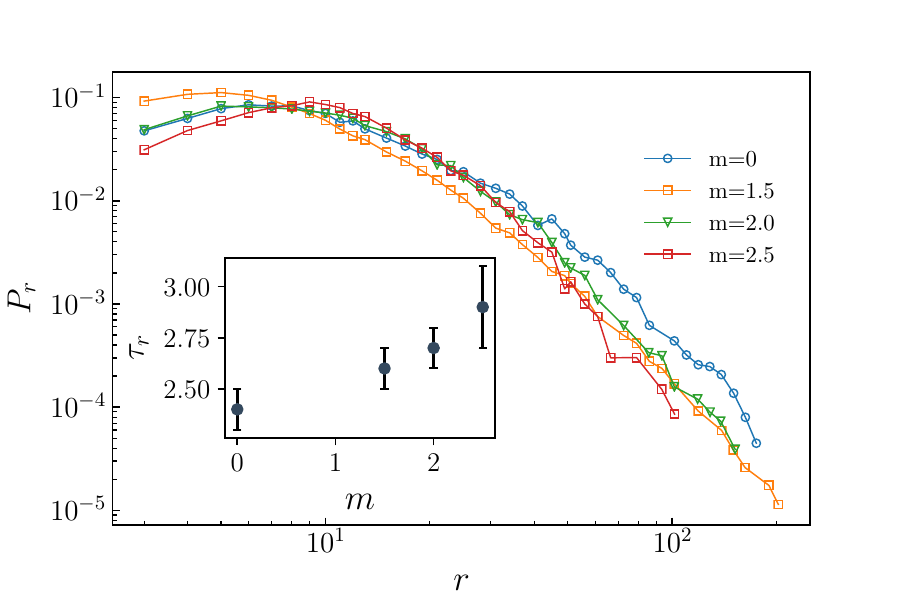}
		\caption{}
		\label{C143}
	\end{subfigure}
	\caption{(a) Mean logarithm of loop length $\left\langle log l\right\rangle $ in terms of the mean logarithm of gyration radius $\left\langle log l\right\rangle $ for various amounts of $m$. Inset shows the fractal dimension of the small (large) scales $\gamma_{lr}^{(1)}$ ($\gamma_{lr}^{(2)}$) in terms of the $m$. Log-log plot of the distribution function of (b) loop length (c) gyration radius, with the insets showing the corresponding exponents in terms of the amount of sugar $m$.}
\end{figure*}
	
An important inquiry pertains to the fractal dimension of level lines, which are transformed into loops in the CLE. If conformal invariance (which is not tested in this study) holds, the fractal dimension $\gamma_{lr}$ of the interfaces can be utilized as a representation of the universality class. Important examples are the fractal dimension of Fortuin-Kasteleyn clusters of the critical Ising model ($\gamma_{lr}=\frac{11}{8}$)~\cite{janke2004geometrical,stella1989scaling,janke2005fractal}, level lines of Gaussian free fields $\gamma_{lr}=\frac{3}{2}$~\cite{cheraghalizadeh2018gaussian2,cheraghalizadeh2018gaussian}, interfaces of 2D percolation theory ($\gamma_{lr}=\frac{7}{4}$)~\cite{tizdast2020dynamical,duplantier2000conformally}, and the frontiers of avalanches in sandpiles ($\gamma_{lr}=\frac{5}{4}$)~\cite{najafi2012avalanche,cheraghalizadeh2017mapping}, for a good review see~\cite{najafi2021some}. Figure \ref{C123} shows the fractal dimension of the loops dried coffee pattern (with and without sugar). In this figure, we used the Eq.~\ref{Eq:FD}, i.e. we plot $\left\langle {\log l} \right\rangle$ in terms of $\left\langle {\log r} \right\rangle$ for different masses of added sugar, the slope of which is the fractal dimension. We observe a cross-over between two distinct spatial regimes where the fractal dimension in small scales is denoted as $\gamma_{lr}^{(1)},$ which differs from the fractal dimensions in larger scales, denoted as $\gamma_{lr}^{(2)}$. As indicated in the inset, the fractal dimensions exhibit negligible dependence on the quantity of sugar used, with $\gamma_{lr}^{(1)} = (1.05 \pm 0.03)$ and $\gamma_{lr}^{(2)} = (1.46 \pm 0.05)$. These results indicate that on small scales, the level lines are not fractal and behave like linear objects. However, they tend to exhibit relatively dense fractal characteristics on larger scales, which is consistent with the behavior of GFFs. To be more precise we have calculated the probability distribution function of the loops, shown in Fig.~\ref{C133}, which confirms that it is power-law is according to Eq.~\ref{Eq:distribution} for large $l$ and $r$ values (note that for small scales these functions are constant, consistent with the observation for the fractal dimension). The exponents should be compared with the exponents of the GFFs, for which with $\tau_l^{\text{GFF}}=\frac{7}{3}\approx 2.33$, and $\tau_r^{\text{GFF}}=3$~\cite{cheraghalizadeh2018gaussian2}. The insets display these exponents as a function of $m$. Our observations reveal that, for sufficiently large $m$ values, these exponents tend to converge towards those of the GFF model, while for the small sugar amounts they are different.
	
\begin{figure*}
		\begin{subfigure}{0.34\textwidth}\includegraphics[width=\textwidth]{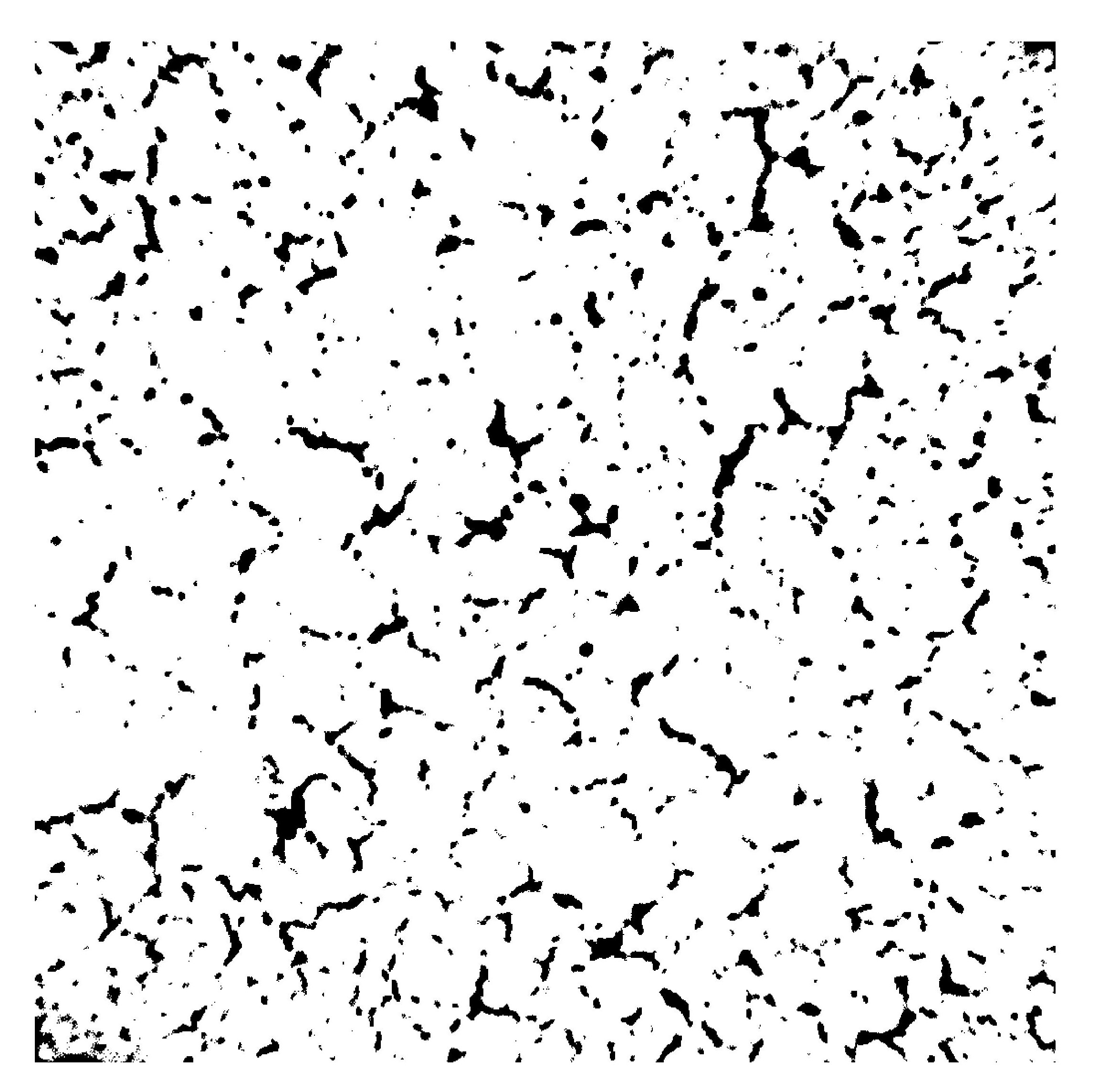}
			\caption{}
			\label{Fig:MA1}
		\end{subfigure}
		\begin{subfigure}{0.5\textwidth}\includegraphics[width=\textwidth]{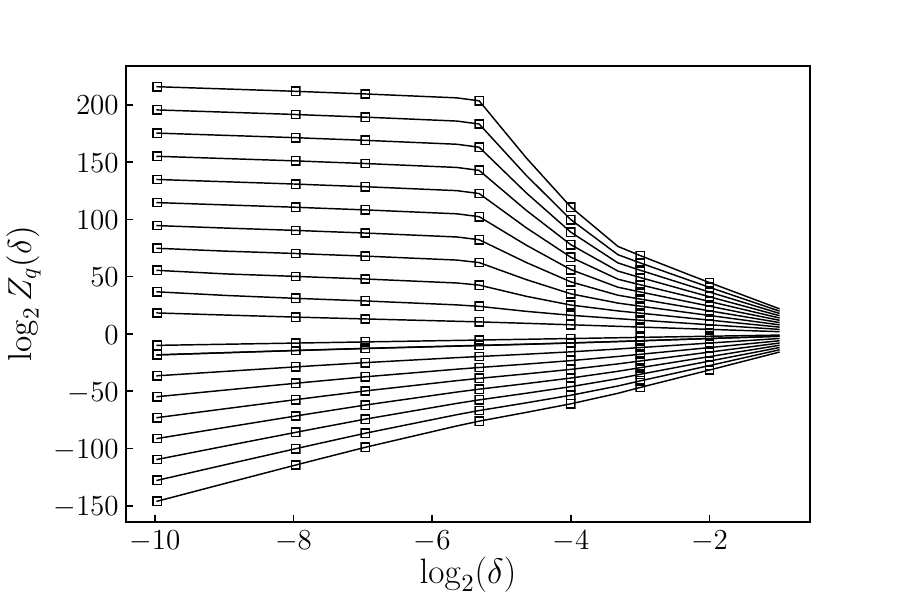}
			\caption{}
			\label{Fig:MA2}
		\end{subfigure}
		\begin{subfigure}{0.48\textwidth}\includegraphics[width=\textwidth]{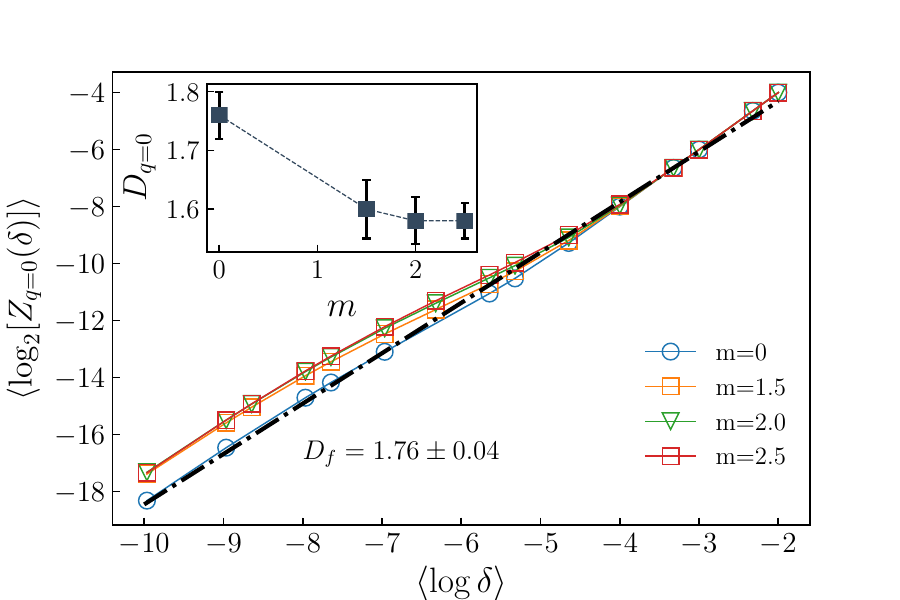}
			\caption{}
			\label{Fig:MA3}
		\end{subfigure}
		\begin{subfigure}{0.48\textwidth}\includegraphics[width=\textwidth]{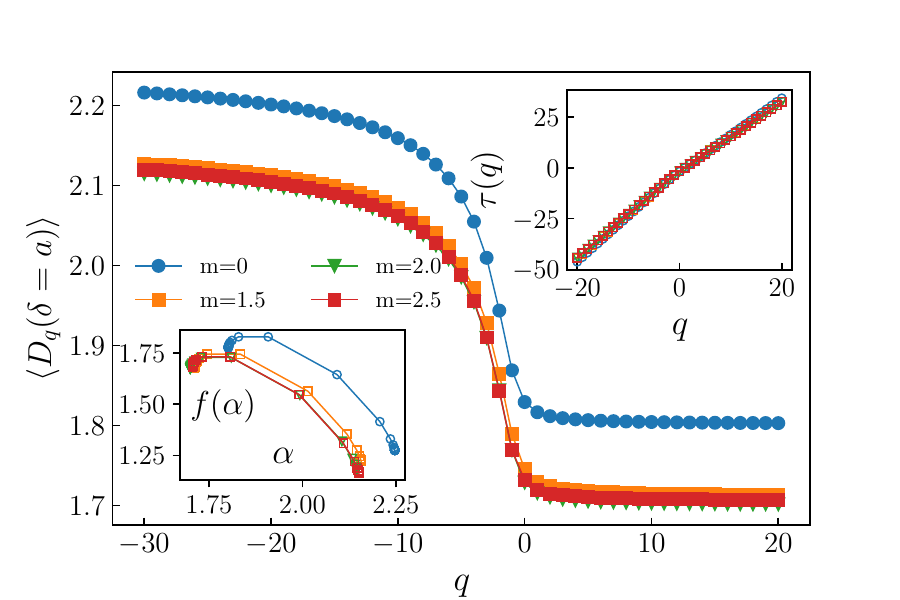}
			\caption{}
			\label{Fig:MA4}
		\end{subfigure}
		\caption{(a) Binary image of dried coffee pattern. (b) Logarithm of the partition function in terms of the logarithm of box liner size $\log \delta$ for $q\in [-20,20]$ with increment 2 ($q$ increases from top to bottom). (c) logarithm of the partition function per $q = 0$ in terms of the $\log \delta$. The inset shows the mass fractal dimension $D_{q=0}$ in terms of the sugar amount $m$. (d) fractal dimension per fixed $\delta$ in terms of $q$ for various $m$ values. Upper inset: $\tau(q)$ in terms of $q$. Lower inset: the exponent spectrum $f(\alpha)$ in terms of $\alpha$.  moment torque diagram in terms of average .}
	\end{figure*}

For our multifractal analysis, we converted the photos from grayscale to binary photos (see Fig.~\ref{Fig:MA1}), so that the samples to be used in MA are $2000 \times 2000$ matrices which are suitable for applying the methodology described in SEC.~\ref{SEC:MA}. Figure~\ref{Fig:MA2} shows $\log_2 Z_q(\delta)$ in terms of $\log_2(\delta)$ (eight $\delta$ values were considered) for various moments $-20\le q\le 20$. Note that, $\tau_q$ is obtained by going to $\delta\rightarrow 0$. Figures~\ref{Fig:MA3} and~\ref{Fig:MA4} show the results for $D_f$ (Eq.~\ref{Eq:FDMA}) and $D_q$ (Eq.\ref{eq73}) respectively. The resulting fractal dimension is shown in the inset. Notably, we observe a sharp decrease in $D_f$ from $1.76\pm 0.04$ to $1.6\pm 0.05$ as we shift from $m=0$ to non-zero $m$ values. This decrease can be attributed to the increased hydrophilicity of the droplet, leading to the emergence of sparser spatial patterns. This trend aligns with the findings of the contact angle statistics. We obtain the multifractal spectrum using Eq.~\ref{Eq:Legendre}, where $f(\alpha)$ represents the spectrum of the fractal dimension. The inset of Fig.~\ref{Fig:MA4} displays this function, which exhibits a similar trend: as $m$ increases, the peak point (i.e., the average $\tau_q$) shifts towards the left, indicating a reduction, thus corroborating the findings discussed earlier. Especially, the peak amount of $f(\alpha_{\text{peak}})$ coincide with the mass fractal dimension $D_f$ that we have found above. Note that the width of the spectrum is not $m$-dependent. A same phenomena is seen for the $q$-fractal dimensions $D_q$, i.e. they decrease as $m$ increases (main panel of Fig.~\ref{Fig:MA4}).

	\section{conclusion}
	In this paper, we statistically analyzed the drying pattern of coffee droplets with and without sugar on the mica sheets. The amount of sugar is controlled by the sugar mass $m$. We discussed various aspects of the drying dynamics, including the evaporation dynamics as well as the Marangoni effect. It was found that the amount of sugar present affects the resulting crack patterns. In this study, we observed the well-known phenomenon of coffee ring formation, where coffee particles accumulate at the edge due to self-pinning. We analyzed the contact angle as a function of $m$, and found that an increase in sugar mass leads to a decrease in the contact angle. Our assertion is that the presence of sugar enhances the hydrophilicity of coffee droplets, resulting in a more even distribution of dried coffee particles for non-zero $m$ values, as compared to the case where $m$ equals zero. 

By mapping the dried patterns for rough surfaces, we constructed a loop ensemble by cutting the samples from a threshold. By analyzing the scaling relation between the loop length and the gyration radius of loops, we numerically estimated the loop fractal dimension, as well as the exponents of the distribution function of loop length and the gyration radius. It was demonstrated that the exponents are in agreement with the Gaussian free fields (GFF) for sufficiently large $m$ values.  The hyper-scaling relation
	\begin{equation}
		\gamma_{lr}=\frac{\tau_r-1}{\tau_l-1}
	\end{equation}
	is valid for all $m$ values.
	
	In the final section of the paper, we utilized multifractal analysis (MA) to study the mass of the system. The methodology for this analysis is detailed in Section~\ref{SEC:MA}. Our findings indicate that the system exhibits multifractality, with a spectrum of fractal dimensions that is dependent on the quantity of sugar present. It was found that the average fractal dimension, represented by the peak point of the spectrum, is dependent on $m$: as $m$ increases, the average fractal dimension decreases. The results align with the contact angle analysis, which indicates that an increase in $m$ leads to a more dispersed dried coffee pattern. For $m=0$, the mass fractal dimension $D_{q=0}$ is shown numerically to be $1.76\pm 0.04$, while it drops to $1.60\pm 0.05$ for non-zero $m$ values, showing that the effect of sugar is considerable.

	
	
\section{acknowledgement}
We acknowledge Dr. Y. Azizi-Kalandaragh for the discussions in the experimental part of this work.

\bibliography{refs}

\begin{thebibliography}{69}%
\makeatletter
\providecommand \@ifxundefined [1]{%
 \@ifx{#1\undefined}
}%
\providecommand \@ifnum [1]{%
 \ifnum #1\expandafter \@firstoftwo
 \else \expandafter \@secondoftwo
 \fi
}%
\providecommand \@ifx [1]{%
 \ifx #1\expandafter \@firstoftwo
 \else \expandafter \@secondoftwo
 \fi
}%
\providecommand \natexlab [1]{#1}%
\providecommand \enquote  [1]{``#1''}%
\providecommand \bibnamefont  [1]{#1}%
\providecommand \bibfnamefont [1]{#1}%
\providecommand \citenamefont [1]{#1}%
\providecommand \href@noop [0]{\@secondoftwo}%
\providecommand \href [0]{\begingroup \@sanitize@url \@href}%
\providecommand \@href[1]{\@@startlink{#1}\@@href}%
\providecommand \@@href[1]{\endgroup#1\@@endlink}%
\providecommand \@sanitize@url [0]{\catcode `\\12\catcode `\$12\catcode
  `\&12\catcode `\#12\catcode `\^12\catcode `\_12\catcode `\%12\relax}%
\providecommand \@@startlink[1]{}%
\providecommand \@@endlink[0]{}%
\providecommand \url  [0]{\begingroup\@sanitize@url \@url }%
\providecommand \@url [1]{\endgroup\@href {#1}{\urlprefix }}%
\providecommand \urlprefix  [0]{URL }%
\providecommand \Eprint [0]{\href }%
\providecommand \doibase [0]{https://doi.org/}%
\providecommand \selectlanguage [0]{\@gobble}%
\providecommand \bibinfo  [0]{\@secondoftwo}%
\providecommand \bibfield  [0]{\@secondoftwo}%
\providecommand \translation [1]{[#1]}%
\providecommand \BibitemOpen [0]{}%
\providecommand \bibitemStop [0]{}%
\providecommand \bibitemNoStop [0]{.\EOS\space}%
\providecommand \EOS [0]{\spacefactor3000\relax}%
\providecommand \BibitemShut  [1]{\csname bibitem#1\endcsname}%
\let\auto@bib@innerbib\@empty
\bibitem [{\citenamefont {Saramito}(2016)}]{saramito2016complex}%
  \BibitemOpen
  \bibfield  {author} {\bibinfo {author} {\bibfnamefont {P.}~\bibnamefont
  {Saramito}},\ }\href@noop {} {\emph {\bibinfo {title} {Complex fluids}}}\
  (\bibinfo  {publisher} {Springer},\ \bibinfo {year} {2016})\BibitemShut
  {NoStop}%
\bibitem [{\citenamefont {Sefiane}(2014)}]{sefiane2014patterns}%
  \BibitemOpen
  \bibfield  {author} {\bibinfo {author} {\bibfnamefont {K.}~\bibnamefont
  {Sefiane}},\ }\href@noop {} {\bibfield  {journal} {\bibinfo  {journal}
  {Advances in colloid and interface science}\ }\textbf {\bibinfo {volume}
  {206}},\ \bibinfo {pages} {372} (\bibinfo {year} {2014})}\BibitemShut
  {NoStop}%
\bibitem [{\citenamefont {Deegan}\ \emph {et~al.}(1997)\citenamefont {Deegan},
  \citenamefont {Bakajin}, \citenamefont {Dupont}, \citenamefont {Huber},
  \citenamefont {Nagel},\ and\ \citenamefont {Witten}}]{deegan1997capillary}%
  \BibitemOpen
  \bibfield  {author} {\bibinfo {author} {\bibfnamefont {R.~D.}\ \bibnamefont
  {Deegan}}, \bibinfo {author} {\bibfnamefont {O.}~\bibnamefont {Bakajin}},
  \bibinfo {author} {\bibfnamefont {T.~F.}\ \bibnamefont {Dupont}}, \bibinfo
  {author} {\bibfnamefont {G.}~\bibnamefont {Huber}}, \bibinfo {author}
  {\bibfnamefont {S.~R.}\ \bibnamefont {Nagel}},\ and\ \bibinfo {author}
  {\bibfnamefont {T.~A.}\ \bibnamefont {Witten}},\ }\href@noop {} {\bibfield
  {journal} {\bibinfo  {journal} {Nature}\ }\textbf {\bibinfo {volume} {389}},\
  \bibinfo {pages} {827} (\bibinfo {year} {1997})}\BibitemShut {NoStop}%
\bibitem [{\citenamefont {Deegan}(2000)}]{deegan2000pattern}%
  \BibitemOpen
  \bibfield  {author} {\bibinfo {author} {\bibfnamefont {R.~D.}\ \bibnamefont
  {Deegan}},\ }\href@noop {} {\bibfield  {journal} {\bibinfo  {journal}
  {Physical review E}\ }\textbf {\bibinfo {volume} {61}},\ \bibinfo {pages}
  {475} (\bibinfo {year} {2000})}\BibitemShut {NoStop}%
\bibitem [{\citenamefont {Zeid}\ and\ \citenamefont
  {Brutin}(2013)}]{zeid2013influence}%
  \BibitemOpen
  \bibfield  {author} {\bibinfo {author} {\bibfnamefont {W.~B.}\ \bibnamefont
  {Zeid}}\ and\ \bibinfo {author} {\bibfnamefont {D.}~\bibnamefont {Brutin}},\
  }\href@noop {} {\bibfield  {journal} {\bibinfo  {journal} {Colloids and
  Surfaces A: Physicochemical and Engineering Aspects}\ }\textbf {\bibinfo
  {volume} {430}},\ \bibinfo {pages} {1} (\bibinfo {year} {2013})}\BibitemShut
  {NoStop}%
\bibitem [{\citenamefont {Chen}\ \emph
  {et~al.}(2016{\natexlab{a}})\citenamefont {Chen}, \citenamefont {Zhang},
  \citenamefont {Zang},\ and\ \citenamefont {Shen}}]{chen2016blood}%
  \BibitemOpen
  \bibfield  {author} {\bibinfo {author} {\bibfnamefont {R.}~\bibnamefont
  {Chen}}, \bibinfo {author} {\bibfnamefont {L.}~\bibnamefont {Zhang}},
  \bibinfo {author} {\bibfnamefont {D.}~\bibnamefont {Zang}},\ and\ \bibinfo
  {author} {\bibfnamefont {W.}~\bibnamefont {Shen}},\ }\href@noop {} {\bibfield
   {journal} {\bibinfo  {journal} {Advances in colloid and interface science}\
  }\textbf {\bibinfo {volume} {231}},\ \bibinfo {pages} {1} (\bibinfo {year}
  {2016}{\natexlab{a}})}\BibitemShut {NoStop}%
\bibitem [{\citenamefont {Chen}\ \emph
  {et~al.}(2016{\natexlab{b}})\citenamefont {Chen}, \citenamefont {Zhang},
  \citenamefont {Zang},\ and\ \citenamefont {Shen}}]{chen2016wetting}%
  \BibitemOpen
  \bibfield  {author} {\bibinfo {author} {\bibfnamefont {R.}~\bibnamefont
  {Chen}}, \bibinfo {author} {\bibfnamefont {L.}~\bibnamefont {Zhang}},
  \bibinfo {author} {\bibfnamefont {D.}~\bibnamefont {Zang}},\ and\ \bibinfo
  {author} {\bibfnamefont {W.}~\bibnamefont {Shen}},\ }\href@noop {} {\bibfield
   {journal} {\bibinfo  {journal} {Advances in Colloid Science}\ ,\ \bibinfo
  {pages} {3}} (\bibinfo {year} {2016}{\natexlab{b}})}\BibitemShut {NoStop}%
\bibitem [{\citenamefont {Hertaeg}\ \emph {et~al.}(2021)\citenamefont
  {Hertaeg}, \citenamefont {Tabor}, \citenamefont {Routh},\ and\ \citenamefont
  {Garnier}}]{hertaeg2021pattern}%
  \BibitemOpen
  \bibfield  {author} {\bibinfo {author} {\bibfnamefont {M.~J.}\ \bibnamefont
  {Hertaeg}}, \bibinfo {author} {\bibfnamefont {R.~F.}\ \bibnamefont {Tabor}},
  \bibinfo {author} {\bibfnamefont {A.~F.}\ \bibnamefont {Routh}},\ and\
  \bibinfo {author} {\bibfnamefont {G.}~\bibnamefont {Garnier}},\ }\href@noop
  {} {\bibfield  {journal} {\bibinfo  {journal} {Philosophical Transactions of
  the Royal Society A}\ }\textbf {\bibinfo {volume} {379}},\ \bibinfo {pages}
  {20200391} (\bibinfo {year} {2021})}\BibitemShut {NoStop}%
\bibitem [{\citenamefont {Shen}\ \emph {et~al.}(2010)\citenamefont {Shen},
  \citenamefont {Ho},\ and\ \citenamefont {Wong}}]{shen2010minimal}%
  \BibitemOpen
  \bibfield  {author} {\bibinfo {author} {\bibfnamefont {X.}~\bibnamefont
  {Shen}}, \bibinfo {author} {\bibfnamefont {C.-M.}\ \bibnamefont {Ho}},\ and\
  \bibinfo {author} {\bibfnamefont {T.-S.}\ \bibnamefont {Wong}},\ }\href@noop
  {} {\bibfield  {journal} {\bibinfo  {journal} {The Journal of Physical
  Chemistry B}\ }\textbf {\bibinfo {volume} {114}},\ \bibinfo {pages} {5269}
  (\bibinfo {year} {2010})}\BibitemShut {NoStop}%
\bibitem [{\citenamefont {Ristenpart}\ \emph {et~al.}(2007)\citenamefont
  {Ristenpart}, \citenamefont {Kim}, \citenamefont {Domingues}, \citenamefont
  {Wan},\ and\ \citenamefont {Stone}}]{ristenpart2007influence}%
  \BibitemOpen
  \bibfield  {author} {\bibinfo {author} {\bibfnamefont {W.}~\bibnamefont
  {Ristenpart}}, \bibinfo {author} {\bibfnamefont {P.}~\bibnamefont {Kim}},
  \bibinfo {author} {\bibfnamefont {C.}~\bibnamefont {Domingues}}, \bibinfo
  {author} {\bibfnamefont {J.}~\bibnamefont {Wan}},\ and\ \bibinfo {author}
  {\bibfnamefont {H.~A.}\ \bibnamefont {Stone}},\ }\href@noop {} {\bibfield
  {journal} {\bibinfo  {journal} {Physical review letters}\ }\textbf {\bibinfo
  {volume} {99}},\ \bibinfo {pages} {234502} (\bibinfo {year}
  {2007})}\BibitemShut {NoStop}%
\bibitem [{\citenamefont {Brutin}(2013)}]{brutin2013influence}%
  \BibitemOpen
  \bibfield  {author} {\bibinfo {author} {\bibfnamefont {D.}~\bibnamefont
  {Brutin}},\ }\href@noop {} {\bibfield  {journal} {\bibinfo  {journal}
  {Colloids and Surfaces A: Physicochemical and Engineering Aspects}\ }\textbf
  {\bibinfo {volume} {429}},\ \bibinfo {pages} {112} (\bibinfo {year}
  {2013})}\BibitemShut {NoStop}%
\bibitem [{\citenamefont {Bahmani}\ \emph {et~al.}(2017)\citenamefont
  {Bahmani}, \citenamefont {Neysari},\ and\ \citenamefont
  {Maleki}}]{bahmani2017study}%
  \BibitemOpen
  \bibfield  {author} {\bibinfo {author} {\bibfnamefont {L.}~\bibnamefont
  {Bahmani}}, \bibinfo {author} {\bibfnamefont {M.}~\bibnamefont {Neysari}},\
  and\ \bibinfo {author} {\bibfnamefont {M.}~\bibnamefont {Maleki}},\
  }\href@noop {} {\bibfield  {journal} {\bibinfo  {journal} {Colloids and
  Surfaces A: Physicochemical and Engineering Aspects}\ }\textbf {\bibinfo
  {volume} {513}},\ \bibinfo {pages} {66} (\bibinfo {year} {2017})}\BibitemShut
  {NoStop}%
\bibitem [{\citenamefont {MUNIANDY}(2013)}]{muniandy2013non}%
  \BibitemOpen
  \bibfield  {author} {\bibinfo {author} {\bibfnamefont {K.}~\bibnamefont
  {MUNIANDY}},\ }\href@noop {} {\bibfield  {journal} {\bibinfo  {journal}
  {Dissertation submitted in partial fulfillment of the requirements for the
  Bachelor of Engineering (Hons) (Chemical Engineering)}\ } (\bibinfo {year}
  {2013})}\BibitemShut {NoStop}%
\bibitem [{\citenamefont {Brutin}\ \emph {et~al.}(2011)\citenamefont {Brutin},
  \citenamefont {Sobac}, \citenamefont {Loquet},\ and\ \citenamefont
  {Sampol}}]{brutin2011pattern}%
  \BibitemOpen
  \bibfield  {author} {\bibinfo {author} {\bibfnamefont {D.}~\bibnamefont
  {Brutin}}, \bibinfo {author} {\bibfnamefont {B.}~\bibnamefont {Sobac}},
  \bibinfo {author} {\bibfnamefont {B.}~\bibnamefont {Loquet}},\ and\ \bibinfo
  {author} {\bibfnamefont {J.}~\bibnamefont {Sampol}},\ }\href@noop {}
  {\bibfield  {journal} {\bibinfo  {journal} {Journal of fluid mechanics}\
  }\textbf {\bibinfo {volume} {667}},\ \bibinfo {pages} {85} (\bibinfo {year}
  {2011})}\BibitemShut {NoStop}%
\bibitem [{\citenamefont {Zeid}\ \emph {et~al.}(2013)\citenamefont {Zeid},
  \citenamefont {Vicente},\ and\ \citenamefont {Brutin}}]{zeid2013influence2}%
  \BibitemOpen
  \bibfield  {author} {\bibinfo {author} {\bibfnamefont {W.~B.}\ \bibnamefont
  {Zeid}}, \bibinfo {author} {\bibfnamefont {J.}~\bibnamefont {Vicente}},\ and\
  \bibinfo {author} {\bibfnamefont {D.}~\bibnamefont {Brutin}},\ }\href@noop {}
  {\bibfield  {journal} {\bibinfo  {journal} {Colloids and Surfaces A:
  Physicochemical and Engineering Aspects}\ }\textbf {\bibinfo {volume}
  {432}},\ \bibinfo {pages} {139} (\bibinfo {year} {2013})}\BibitemShut
  {NoStop}%
\bibitem [{\citenamefont {Pal}\ \emph {et~al.}(2020)\citenamefont {Pal},
  \citenamefont {Gope}, \citenamefont {Obayemi},\ and\ \citenamefont
  {Iannacchione}}]{pal2020concentration}%
  \BibitemOpen
  \bibfield  {author} {\bibinfo {author} {\bibfnamefont {A.}~\bibnamefont
  {Pal}}, \bibinfo {author} {\bibfnamefont {A.}~\bibnamefont {Gope}}, \bibinfo
  {author} {\bibfnamefont {J.~D.}\ \bibnamefont {Obayemi}},\ and\ \bibinfo
  {author} {\bibfnamefont {G.~S.}\ \bibnamefont {Iannacchione}},\ }\href@noop
  {} {\bibfield  {journal} {\bibinfo  {journal} {Scientific reports}\ }\textbf
  {\bibinfo {volume} {10}},\ \bibinfo {pages} {1} (\bibinfo {year}
  {2020})}\BibitemShut {NoStop}%
\bibitem [{\citenamefont {Mondal}\ \emph {et~al.}(2023)\citenamefont {Mondal},
  \citenamefont {Lama},\ and\ \citenamefont {Sahu}}]{mondal2023physics}%
  \BibitemOpen
  \bibfield  {author} {\bibinfo {author} {\bibfnamefont {R.}~\bibnamefont
  {Mondal}}, \bibinfo {author} {\bibfnamefont {H.}~\bibnamefont {Lama}},\ and\
  \bibinfo {author} {\bibfnamefont {K.~C.}\ \bibnamefont {Sahu}},\ }\href@noop
  {} {\bibfield  {journal} {\bibinfo  {journal} {Physics of Fluids}\ }\textbf
  {\bibinfo {volume} {35}} (\bibinfo {year} {2023})}\BibitemShut {NoStop}%
\bibitem [{\citenamefont {Deegan}\ \emph {et~al.}(2000)\citenamefont {Deegan},
  \citenamefont {Bakajin}, \citenamefont {Dupont}, \citenamefont {Huber},
  \citenamefont {Nagel},\ and\ \citenamefont {Witten}}]{deegan2000contact}%
  \BibitemOpen
  \bibfield  {author} {\bibinfo {author} {\bibfnamefont {R.~D.}\ \bibnamefont
  {Deegan}}, \bibinfo {author} {\bibfnamefont {O.}~\bibnamefont {Bakajin}},
  \bibinfo {author} {\bibfnamefont {T.~F.}\ \bibnamefont {Dupont}}, \bibinfo
  {author} {\bibfnamefont {G.}~\bibnamefont {Huber}}, \bibinfo {author}
  {\bibfnamefont {S.~R.}\ \bibnamefont {Nagel}},\ and\ \bibinfo {author}
  {\bibfnamefont {T.~A.}\ \bibnamefont {Witten}},\ }\href@noop {} {\bibfield
  {journal} {\bibinfo  {journal} {Physical review E}\ }\textbf {\bibinfo
  {volume} {62}},\ \bibinfo {pages} {756} (\bibinfo {year} {2000})}\BibitemShut
  {NoStop}%
\bibitem [{\citenamefont {Tekin}\ \emph {et~al.}(2008)\citenamefont {Tekin},
  \citenamefont {Smith},\ and\ \citenamefont {Schubert}}]{tekin2008inkjet}%
  \BibitemOpen
  \bibfield  {author} {\bibinfo {author} {\bibfnamefont {E.}~\bibnamefont
  {Tekin}}, \bibinfo {author} {\bibfnamefont {P.~J.}\ \bibnamefont {Smith}},\
  and\ \bibinfo {author} {\bibfnamefont {U.~S.}\ \bibnamefont {Schubert}},\
  }\href@noop {} {\bibfield  {journal} {\bibinfo  {journal} {Soft Matter}\
  }\textbf {\bibinfo {volume} {4}},\ \bibinfo {pages} {703} (\bibinfo {year}
  {2008})}\BibitemShut {NoStop}%
\bibitem [{\citenamefont {Derby}(2010)}]{derby2010inkjet}%
  \BibitemOpen
  \bibfield  {author} {\bibinfo {author} {\bibfnamefont {B.}~\bibnamefont
  {Derby}},\ }\href@noop {} {\bibfield  {journal} {\bibinfo  {journal} {Annual
  Review of Materials Research}\ }\textbf {\bibinfo {volume} {40}},\ \bibinfo
  {pages} {395} (\bibinfo {year} {2010})}\BibitemShut {NoStop}%
\bibitem [{\citenamefont {Fischer}(2002)}]{fischer2002particle}%
  \BibitemOpen
  \bibfield  {author} {\bibinfo {author} {\bibfnamefont {B.~J.}\ \bibnamefont
  {Fischer}},\ }\href@noop {} {\bibfield  {journal} {\bibinfo  {journal}
  {Langmuir}\ }\textbf {\bibinfo {volume} {18}},\ \bibinfo {pages} {60}
  (\bibinfo {year} {2002})}\BibitemShut {NoStop}%
\bibitem [{\citenamefont {Popov}(2005)}]{popov2005evaporative}%
  \BibitemOpen
  \bibfield  {author} {\bibinfo {author} {\bibfnamefont {Y.~O.}\ \bibnamefont
  {Popov}},\ }\href@noop {} {\bibfield  {journal} {\bibinfo  {journal}
  {Physical Review E}\ }\textbf {\bibinfo {volume} {71}},\ \bibinfo {pages}
  {036313} (\bibinfo {year} {2005})}\BibitemShut {NoStop}%
\bibitem [{\citenamefont {Xu}\ \emph {et~al.}(2006)\citenamefont {Xu},
  \citenamefont {Xia}, \citenamefont {Hong}, \citenamefont {Lin}, \citenamefont
  {Qiu},\ and\ \citenamefont {Yang}}]{xu2006self}%
  \BibitemOpen
  \bibfield  {author} {\bibinfo {author} {\bibfnamefont {J.}~\bibnamefont
  {Xu}}, \bibinfo {author} {\bibfnamefont {J.}~\bibnamefont {Xia}}, \bibinfo
  {author} {\bibfnamefont {S.~W.}\ \bibnamefont {Hong}}, \bibinfo {author}
  {\bibfnamefont {Z.}~\bibnamefont {Lin}}, \bibinfo {author} {\bibfnamefont
  {F.}~\bibnamefont {Qiu}},\ and\ \bibinfo {author} {\bibfnamefont
  {Y.}~\bibnamefont {Yang}},\ }\href@noop {} {\bibfield  {journal} {\bibinfo
  {journal} {Physical review letters}\ }\textbf {\bibinfo {volume} {96}},\
  \bibinfo {pages} {066104} (\bibinfo {year} {2006})}\BibitemShut {NoStop}%
\bibitem [{\citenamefont {Hu}\ and\ \citenamefont
  {Larson}(2006)}]{hu2006marangoni}%
  \BibitemOpen
  \bibfield  {author} {\bibinfo {author} {\bibfnamefont {H.}~\bibnamefont
  {Hu}}\ and\ \bibinfo {author} {\bibfnamefont {R.~G.}\ \bibnamefont
  {Larson}},\ }\href@noop {} {\bibfield  {journal} {\bibinfo  {journal} {The
  Journal of Physical Chemistry B}\ }\textbf {\bibinfo {volume} {110}},\
  \bibinfo {pages} {7090} (\bibinfo {year} {2006})}\BibitemShut {NoStop}%
\bibitem [{\citenamefont {Park}\ and\ \citenamefont
  {Moon}(2006)}]{park2006control}%
  \BibitemOpen
  \bibfield  {author} {\bibinfo {author} {\bibfnamefont {J.}~\bibnamefont
  {Park}}\ and\ \bibinfo {author} {\bibfnamefont {J.}~\bibnamefont {Moon}},\
  }\href@noop {} {\bibfield  {journal} {\bibinfo  {journal} {Langmuir}\
  }\textbf {\bibinfo {volume} {22}},\ \bibinfo {pages} {3506} (\bibinfo {year}
  {2006})}\BibitemShut {NoStop}%
\bibitem [{\citenamefont {Dugas}\ \emph {et~al.}(2005)\citenamefont {Dugas},
  \citenamefont {Broutin},\ and\ \citenamefont
  {Souteyrand}}]{dugas2005droplet}%
  \BibitemOpen
  \bibfield  {author} {\bibinfo {author} {\bibfnamefont {V.}~\bibnamefont
  {Dugas}}, \bibinfo {author} {\bibfnamefont {J.}~\bibnamefont {Broutin}},\
  and\ \bibinfo {author} {\bibfnamefont {E.}~\bibnamefont {Souteyrand}},\
  }\href@noop {} {\bibfield  {journal} {\bibinfo  {journal} {Langmuir}\
  }\textbf {\bibinfo {volume} {21}},\ \bibinfo {pages} {9130} (\bibinfo {year}
  {2005})}\BibitemShut {NoStop}%
\bibitem [{\citenamefont {Yunker}\ \emph {et~al.}(2011)\citenamefont {Yunker},
  \citenamefont {Still}, \citenamefont {Lohr},\ and\ \citenamefont
  {Yodh}}]{yunker2011suppression}%
  \BibitemOpen
  \bibfield  {author} {\bibinfo {author} {\bibfnamefont {P.~J.}\ \bibnamefont
  {Yunker}}, \bibinfo {author} {\bibfnamefont {T.}~\bibnamefont {Still}},
  \bibinfo {author} {\bibfnamefont {M.~A.}\ \bibnamefont {Lohr}},\ and\
  \bibinfo {author} {\bibfnamefont {A.}~\bibnamefont {Yodh}},\ }\href@noop {}
  {\bibfield  {journal} {\bibinfo  {journal} {nature}\ }\textbf {\bibinfo
  {volume} {476}},\ \bibinfo {pages} {308} (\bibinfo {year}
  {2011})}\BibitemShut {NoStop}%
\bibitem [{\citenamefont {Still}\ \emph {et~al.}(2012)\citenamefont {Still},
  \citenamefont {Yunker},\ and\ \citenamefont {Yodh}}]{still2012surfactant}%
  \BibitemOpen
  \bibfield  {author} {\bibinfo {author} {\bibfnamefont {T.}~\bibnamefont
  {Still}}, \bibinfo {author} {\bibfnamefont {P.~J.}\ \bibnamefont {Yunker}},\
  and\ \bibinfo {author} {\bibfnamefont {A.~G.}\ \bibnamefont {Yodh}},\
  }\href@noop {} {\bibfield  {journal} {\bibinfo  {journal} {Langmuir}\
  }\textbf {\bibinfo {volume} {28}},\ \bibinfo {pages} {4984} (\bibinfo {year}
  {2012})}\BibitemShut {NoStop}%
\bibitem [{\citenamefont {Govor}\ \emph {et~al.}(2004)\citenamefont {Govor},
  \citenamefont {Reiter}, \citenamefont {Bauer},\ and\ \citenamefont
  {Parisi}}]{govor2004nanoparticle}%
  \BibitemOpen
  \bibfield  {author} {\bibinfo {author} {\bibfnamefont {L.}~\bibnamefont
  {Govor}}, \bibinfo {author} {\bibfnamefont {G.}~\bibnamefont {Reiter}},
  \bibinfo {author} {\bibfnamefont {G.}~\bibnamefont {Bauer}},\ and\ \bibinfo
  {author} {\bibfnamefont {J.}~\bibnamefont {Parisi}},\ }\href@noop {}
  {\bibfield  {journal} {\bibinfo  {journal} {Applied physics letters}\
  }\textbf {\bibinfo {volume} {84}},\ \bibinfo {pages} {4774} (\bibinfo {year}
  {2004})}\BibitemShut {NoStop}%
\bibitem [{\citenamefont {Smalyukh}\ \emph {et~al.}(2006)\citenamefont
  {Smalyukh}, \citenamefont {Zribi}, \citenamefont {Butler}, \citenamefont
  {Lavrentovich},\ and\ \citenamefont {Wong}}]{smalyukh2006structure}%
  \BibitemOpen
  \bibfield  {author} {\bibinfo {author} {\bibfnamefont {I.~I.}\ \bibnamefont
  {Smalyukh}}, \bibinfo {author} {\bibfnamefont {O.~V.}\ \bibnamefont {Zribi}},
  \bibinfo {author} {\bibfnamefont {J.~C.}\ \bibnamefont {Butler}}, \bibinfo
  {author} {\bibfnamefont {O.~D.}\ \bibnamefont {Lavrentovich}},\ and\ \bibinfo
  {author} {\bibfnamefont {G.~C.}\ \bibnamefont {Wong}},\ }\href@noop {}
  {\bibfield  {journal} {\bibinfo  {journal} {Physical review letters}\
  }\textbf {\bibinfo {volume} {96}},\ \bibinfo {pages} {177801} (\bibinfo
  {year} {2006})}\BibitemShut {NoStop}%
\bibitem [{\citenamefont {Yen}\ \emph {et~al.}(2018)\citenamefont {Yen},
  \citenamefont {Fu}, \citenamefont {Wei}, \citenamefont {Nayak}, \citenamefont
  {Shi},\ and\ \citenamefont {Lo}}]{yen2018reversing}%
  \BibitemOpen
  \bibfield  {author} {\bibinfo {author} {\bibfnamefont {T.~M.}\ \bibnamefont
  {Yen}}, \bibinfo {author} {\bibfnamefont {X.}~\bibnamefont {Fu}}, \bibinfo
  {author} {\bibfnamefont {T.}~\bibnamefont {Wei}}, \bibinfo {author}
  {\bibfnamefont {R.~U.}\ \bibnamefont {Nayak}}, \bibinfo {author}
  {\bibfnamefont {Y.}~\bibnamefont {Shi}},\ and\ \bibinfo {author}
  {\bibfnamefont {Y.-H.}\ \bibnamefont {Lo}},\ }\href@noop {} {\bibfield
  {journal} {\bibinfo  {journal} {Scientific reports}\ }\textbf {\bibinfo
  {volume} {8}},\ \bibinfo {pages} {1} (\bibinfo {year} {2018})}\BibitemShut
  {NoStop}%
\bibitem [{\citenamefont {Zang}\ \emph {et~al.}(2019)\citenamefont {Zang},
  \citenamefont {Tarafdar}, \citenamefont {Tarasevich}, \citenamefont
  {Choudhury},\ and\ \citenamefont {Dutta}}]{zang2019evaporation}%
  \BibitemOpen
  \bibfield  {author} {\bibinfo {author} {\bibfnamefont {D.}~\bibnamefont
  {Zang}}, \bibinfo {author} {\bibfnamefont {S.}~\bibnamefont {Tarafdar}},
  \bibinfo {author} {\bibfnamefont {Y.~Y.}\ \bibnamefont {Tarasevich}},
  \bibinfo {author} {\bibfnamefont {M.~D.}\ \bibnamefont {Choudhury}},\ and\
  \bibinfo {author} {\bibfnamefont {T.}~\bibnamefont {Dutta}},\ }\href@noop {}
  {\bibfield  {journal} {\bibinfo  {journal} {Physics Reports}\ }\textbf
  {\bibinfo {volume} {804}},\ \bibinfo {pages} {1} (\bibinfo {year}
  {2019})}\BibitemShut {NoStop}%
\bibitem [{\citenamefont {Thampi}\ and\ \citenamefont
  {Basavaraj}(2020)}]{thampi2020beyond}%
  \BibitemOpen
  \bibfield  {author} {\bibinfo {author} {\bibfnamefont {S.~P.}\ \bibnamefont
  {Thampi}}\ and\ \bibinfo {author} {\bibfnamefont {M.~G.}\ \bibnamefont
  {Basavaraj}},\ }\href@noop {} {\bibfield  {journal} {\bibinfo  {journal} {ACS
  omega}\ }\textbf {\bibinfo {volume} {5}},\ \bibinfo {pages} {11262} (\bibinfo
  {year} {2020})}\BibitemShut {NoStop}%
\bibitem [{\citenamefont {Craster}\ and\ \citenamefont
  {Matar}(2009)}]{craster2009dynamics}%
  \BibitemOpen
  \bibfield  {author} {\bibinfo {author} {\bibfnamefont {R.~V.}\ \bibnamefont
  {Craster}}\ and\ \bibinfo {author} {\bibfnamefont {O.~K.}\ \bibnamefont
  {Matar}},\ }\href@noop {} {\bibfield  {journal} {\bibinfo  {journal} {Reviews
  of modern physics}\ }\textbf {\bibinfo {volume} {81}},\ \bibinfo {pages}
  {1131} (\bibinfo {year} {2009})}\BibitemShut {NoStop}%
\bibitem [{\citenamefont {Bonn}\ \emph {et~al.}(2009)\citenamefont {Bonn},
  \citenamefont {Eggers}, \citenamefont {Indekeu}, \citenamefont {Meunier},\
  and\ \citenamefont {Rolley}}]{bonn2009wetting}%
  \BibitemOpen
  \bibfield  {author} {\bibinfo {author} {\bibfnamefont {D.}~\bibnamefont
  {Bonn}}, \bibinfo {author} {\bibfnamefont {J.}~\bibnamefont {Eggers}},
  \bibinfo {author} {\bibfnamefont {J.}~\bibnamefont {Indekeu}}, \bibinfo
  {author} {\bibfnamefont {J.}~\bibnamefont {Meunier}},\ and\ \bibinfo {author}
  {\bibfnamefont {E.}~\bibnamefont {Rolley}},\ }\href@noop {} {\bibfield
  {journal} {\bibinfo  {journal} {Reviews of modern physics}\ }\textbf
  {\bibinfo {volume} {81}},\ \bibinfo {pages} {739} (\bibinfo {year}
  {2009})}\BibitemShut {NoStop}%
\bibitem [{\citenamefont {Han}\ and\ \citenamefont
  {Lin}(2012)}]{han2012learning}%
  \BibitemOpen
  \bibfield  {author} {\bibinfo {author} {\bibfnamefont {W.}~\bibnamefont
  {Han}}\ and\ \bibinfo {author} {\bibfnamefont {Z.}~\bibnamefont {Lin}},\
  }\href@noop {} {\bibfield  {journal} {\bibinfo  {journal} {Angewandte Chemie
  International Edition}\ }\textbf {\bibinfo {volume} {51}},\ \bibinfo {pages}
  {1534} (\bibinfo {year} {2012})}\BibitemShut {NoStop}%
\bibitem [{\citenamefont {Erbil}(2012)}]{erbil2012evaporation}%
  \BibitemOpen
  \bibfield  {author} {\bibinfo {author} {\bibfnamefont {H.~Y.}\ \bibnamefont
  {Erbil}},\ }\href@noop {} {\bibfield  {journal} {\bibinfo  {journal}
  {Advances in colloid and interface science}\ }\textbf {\bibinfo {volume}
  {170}},\ \bibinfo {pages} {67} (\bibinfo {year} {2012})}\BibitemShut
  {NoStop}%
\bibitem [{\citenamefont {Brutin}\ and\ \citenamefont
  {Starov}(2018)}]{brutin2018recent}%
  \BibitemOpen
  \bibfield  {author} {\bibinfo {author} {\bibfnamefont {D.}~\bibnamefont
  {Brutin}}\ and\ \bibinfo {author} {\bibfnamefont {V.}~\bibnamefont
  {Starov}},\ }\href@noop {} {\bibfield  {journal} {\bibinfo  {journal}
  {Chemical Society Reviews}\ }\textbf {\bibinfo {volume} {47}},\ \bibinfo
  {pages} {558} (\bibinfo {year} {2018})}\BibitemShut {NoStop}%
\bibitem [{\citenamefont {Mampallil}\ and\ \citenamefont
  {Eral}(2018)}]{mampallil2018review}%
  \BibitemOpen
  \bibfield  {author} {\bibinfo {author} {\bibfnamefont {D.}~\bibnamefont
  {Mampallil}}\ and\ \bibinfo {author} {\bibfnamefont {H.~B.}\ \bibnamefont
  {Eral}},\ }\href@noop {} {\bibfield  {journal} {\bibinfo  {journal} {Advances
  in colloid and interface science}\ }\textbf {\bibinfo {volume} {252}},\
  \bibinfo {pages} {38} (\bibinfo {year} {2018})}\BibitemShut {NoStop}%
\bibitem [{\citenamefont {Al-Milaji}\ and\ \citenamefont
  {Zhao}(2019)}]{al2019new}%
  \BibitemOpen
  \bibfield  {author} {\bibinfo {author} {\bibfnamefont {K.~N.}\ \bibnamefont
  {Al-Milaji}}\ and\ \bibinfo {author} {\bibfnamefont {H.}~\bibnamefont
  {Zhao}},\ }\href@noop {} {\bibfield  {journal} {\bibinfo  {journal} {The
  Journal of Physical Chemistry C}\ }\textbf {\bibinfo {volume} {123}},\
  \bibinfo {pages} {12029} (\bibinfo {year} {2019})}\BibitemShut {NoStop}%
\bibitem [{\citenamefont {Bhardwaj}\ \emph {et~al.}(2010)\citenamefont
  {Bhardwaj}, \citenamefont {Fang}, \citenamefont {Somasundaran},\ and\
  \citenamefont {Attinger}}]{bhardwaj2010self}%
  \BibitemOpen
  \bibfield  {author} {\bibinfo {author} {\bibfnamefont {R.}~\bibnamefont
  {Bhardwaj}}, \bibinfo {author} {\bibfnamefont {X.}~\bibnamefont {Fang}},
  \bibinfo {author} {\bibfnamefont {P.}~\bibnamefont {Somasundaran}},\ and\
  \bibinfo {author} {\bibfnamefont {D.}~\bibnamefont {Attinger}},\ }\href@noop
  {} {\bibfield  {journal} {\bibinfo  {journal} {Langmuir}\ }\textbf {\bibinfo
  {volume} {26}},\ \bibinfo {pages} {7833} (\bibinfo {year}
  {2010})}\BibitemShut {NoStop}%
\bibitem [{\citenamefont {Bird}\ \emph {et~al.}(1987)\citenamefont {Bird},
  \citenamefont {Curtiss}, \citenamefont {Armstrong},\ and\ \citenamefont
  {Hassager}}]{bird1987dynamics}%
  \BibitemOpen
  \bibfield  {author} {\bibinfo {author} {\bibfnamefont {R.~B.}\ \bibnamefont
  {Bird}}, \bibinfo {author} {\bibfnamefont {C.~F.}\ \bibnamefont {Curtiss}},
  \bibinfo {author} {\bibfnamefont {R.~C.}\ \bibnamefont {Armstrong}},\ and\
  \bibinfo {author} {\bibfnamefont {O.}~\bibnamefont {Hassager}},\ }\href@noop
  {} {\emph {\bibinfo {title} {Dynamics of polymeric liquids, volume 2: Kinetic
  theory}}}\ (\bibinfo  {publisher} {Wiley},\ \bibinfo {year}
  {1987})\BibitemShut {NoStop}%
\bibitem [{\citenamefont {Roch{\'e}}\ \emph {et~al.}(2014)\citenamefont
  {Roch{\'e}}, \citenamefont {Li}, \citenamefont {Griffiths}, \citenamefont
  {Le~Roux}, \citenamefont {Cantat}, \citenamefont {Saint-Jalmes},\ and\
  \citenamefont {Stone}}]{roche2014marangoni}%
  \BibitemOpen
  \bibfield  {author} {\bibinfo {author} {\bibfnamefont {M.}~\bibnamefont
  {Roch{\'e}}}, \bibinfo {author} {\bibfnamefont {Z.}~\bibnamefont {Li}},
  \bibinfo {author} {\bibfnamefont {I.~M.}\ \bibnamefont {Griffiths}}, \bibinfo
  {author} {\bibfnamefont {S.}~\bibnamefont {Le~Roux}}, \bibinfo {author}
  {\bibfnamefont {I.}~\bibnamefont {Cantat}}, \bibinfo {author} {\bibfnamefont
  {A.}~\bibnamefont {Saint-Jalmes}},\ and\ \bibinfo {author} {\bibfnamefont
  {H.~A.}\ \bibnamefont {Stone}},\ }\href@noop {} {\bibfield  {journal}
  {\bibinfo  {journal} {Physical review letters}\ }\textbf {\bibinfo {volume}
  {112}},\ \bibinfo {pages} {208302} (\bibinfo {year} {2014})}\BibitemShut
  {NoStop}%
\bibitem [{\citenamefont {Gaston}(2021)}]{gaston2021marangoni}%
  \BibitemOpen
  \bibfield  {author} {\bibinfo {author} {\bibfnamefont {T.}~\bibnamefont
  {Gaston}},\ }\href@noop {} {\  (\bibinfo {year} {2021})}\BibitemShut
  {NoStop}%
\bibitem [{\citenamefont {Keiser}\ \emph {et~al.}(2017)\citenamefont {Keiser},
  \citenamefont {Bense}, \citenamefont {Colinet}, \citenamefont {Bico},\ and\
  \citenamefont {Reyssat}}]{keiser2017marangoni}%
  \BibitemOpen
  \bibfield  {author} {\bibinfo {author} {\bibfnamefont {L.}~\bibnamefont
  {Keiser}}, \bibinfo {author} {\bibfnamefont {H.}~\bibnamefont {Bense}},
  \bibinfo {author} {\bibfnamefont {P.}~\bibnamefont {Colinet}}, \bibinfo
  {author} {\bibfnamefont {J.}~\bibnamefont {Bico}},\ and\ \bibinfo {author}
  {\bibfnamefont {E.}~\bibnamefont {Reyssat}},\ }\href@noop {} {\bibfield
  {journal} {\bibinfo  {journal} {Physical review letters}\ }\textbf {\bibinfo
  {volume} {118}},\ \bibinfo {pages} {074504} (\bibinfo {year}
  {2017})}\BibitemShut {NoStop}%
\bibitem [{\citenamefont {Biswas}\ \emph {et~al.}(2018)\citenamefont {Biswas},
  \citenamefont {Fantinel}, \citenamefont {Borgman}, \citenamefont {Holtzman},\
  and\ \citenamefont {Goehring}}]{biswas2018drying}%
  \BibitemOpen
  \bibfield  {author} {\bibinfo {author} {\bibfnamefont {S.}~\bibnamefont
  {Biswas}}, \bibinfo {author} {\bibfnamefont {P.}~\bibnamefont {Fantinel}},
  \bibinfo {author} {\bibfnamefont {O.}~\bibnamefont {Borgman}}, \bibinfo
  {author} {\bibfnamefont {R.}~\bibnamefont {Holtzman}},\ and\ \bibinfo
  {author} {\bibfnamefont {L.}~\bibnamefont {Goehring}},\ }\href@noop {}
  {\bibfield  {journal} {\bibinfo  {journal} {Physical Review Fluids}\ }\textbf
  {\bibinfo {volume} {3}},\ \bibinfo {pages} {124307} (\bibinfo {year}
  {2018})}\BibitemShut {NoStop}%
\bibitem [{\citenamefont {Thiele}(2014)}]{thiele2014patterned}%
  \BibitemOpen
  \bibfield  {author} {\bibinfo {author} {\bibfnamefont {U.}~\bibnamefont
  {Thiele}},\ }\href@noop {} {\bibfield  {journal} {\bibinfo  {journal}
  {Advances in colloid and interface science}\ }\textbf {\bibinfo {volume}
  {206}},\ \bibinfo {pages} {399} (\bibinfo {year} {2014})}\BibitemShut
  {NoStop}%
\bibitem [{\citenamefont {Taniguchi}\ \emph {et~al.}(2003)\citenamefont
  {Taniguchi}, \citenamefont {Harada}, \citenamefont {Kojo}, \citenamefont
  {Nakayama},\ and\ \citenamefont {Wakao}}]{taniguchi2003regulatory}%
  \BibitemOpen
  \bibfield  {author} {\bibinfo {author} {\bibfnamefont {M.}~\bibnamefont
  {Taniguchi}}, \bibinfo {author} {\bibfnamefont {M.}~\bibnamefont {Harada}},
  \bibinfo {author} {\bibfnamefont {S.}~\bibnamefont {Kojo}}, \bibinfo {author}
  {\bibfnamefont {T.}~\bibnamefont {Nakayama}},\ and\ \bibinfo {author}
  {\bibfnamefont {H.}~\bibnamefont {Wakao}},\ }\href@noop {} {\bibfield
  {journal} {\bibinfo  {journal} {Annual review of immunology}\ }\textbf
  {\bibinfo {volume} {21}},\ \bibinfo {pages} {483} (\bibinfo {year}
  {2003})}\BibitemShut {NoStop}%
\bibitem [{Note1()}]{Note1}%
  \BibitemOpen
  \bibinfo {note} {ImageJ software bundled with 64-bit Java 8}\BibitemShut
  {NoStop}%
\bibitem [{\citenamefont {Janke}\ and\ \citenamefont
  {Schakel}(2004)}]{janke2004geometrical}%
  \BibitemOpen
  \bibfield  {author} {\bibinfo {author} {\bibfnamefont {W.}~\bibnamefont
  {Janke}}\ and\ \bibinfo {author} {\bibfnamefont {A.~M.}\ \bibnamefont
  {Schakel}},\ }\href@noop {} {\bibfield  {journal} {\bibinfo  {journal}
  {Nuclear Physics B}\ }\textbf {\bibinfo {volume} {700}},\ \bibinfo {pages}
  {385} (\bibinfo {year} {2004})}\BibitemShut {NoStop}%
\bibitem [{\citenamefont {Stella}\ and\ \citenamefont
  {Vanderzande}(1989)}]{stella1989scaling}%
  \BibitemOpen
  \bibfield  {author} {\bibinfo {author} {\bibfnamefont {A.}~\bibnamefont
  {Stella}}\ and\ \bibinfo {author} {\bibfnamefont {C.}~\bibnamefont
  {Vanderzande}},\ }\href@noop {} {\bibfield  {journal} {\bibinfo  {journal}
  {Physical review letters}\ }\textbf {\bibinfo {volume} {62}},\ \bibinfo
  {pages} {1067} (\bibinfo {year} {1989})}\BibitemShut {NoStop}%
\bibitem [{\citenamefont {Janke}\ and\ \citenamefont
  {Schakel}(2005)}]{janke2005fractal}%
  \BibitemOpen
  \bibfield  {author} {\bibinfo {author} {\bibfnamefont {W.}~\bibnamefont
  {Janke}}\ and\ \bibinfo {author} {\bibfnamefont {A.~M.}\ \bibnamefont
  {Schakel}},\ }\href@noop {} {\bibfield  {journal} {\bibinfo  {journal}
  {Physical Review E}\ }\textbf {\bibinfo {volume} {71}},\ \bibinfo {pages}
  {036703} (\bibinfo {year} {2005})}\BibitemShut {NoStop}%
\bibitem [{\citenamefont {Cheraghalizadeh}\ \emph
  {et~al.}(2018{\natexlab{a}})\citenamefont {Cheraghalizadeh}, \citenamefont
  {Najafi},\ and\ \citenamefont
  {Mohammadzadeh}}]{cheraghalizadeh2018gaussian2}%
  \BibitemOpen
  \bibfield  {author} {\bibinfo {author} {\bibfnamefont {J.}~\bibnamefont
  {Cheraghalizadeh}}, \bibinfo {author} {\bibfnamefont {M.~N.}\ \bibnamefont
  {Najafi}},\ and\ \bibinfo {author} {\bibfnamefont {H.}~\bibnamefont
  {Mohammadzadeh}},\ }\href@noop {} {\bibfield  {journal} {\bibinfo  {journal}
  {The European Physical Journal B}\ }\textbf {\bibinfo {volume} {91}},\
  \bibinfo {pages} {1} (\bibinfo {year} {2018}{\natexlab{a}})}\BibitemShut
  {NoStop}%
\bibitem [{\citenamefont {Cheraghalizadeh}\ \emph
  {et~al.}(2018{\natexlab{b}})\citenamefont {Cheraghalizadeh}, \citenamefont
  {Najafi},\ and\ \citenamefont {Mohammadzadeh}}]{cheraghalizadeh2018gaussian}%
  \BibitemOpen
  \bibfield  {author} {\bibinfo {author} {\bibfnamefont {J.}~\bibnamefont
  {Cheraghalizadeh}}, \bibinfo {author} {\bibfnamefont {M.}~\bibnamefont
  {Najafi}},\ and\ \bibinfo {author} {\bibfnamefont {H.}~\bibnamefont
  {Mohammadzadeh}},\ }\href@noop {} {\bibfield  {journal} {\bibinfo  {journal}
  {Journal of Statistical Mechanics: Theory and Experiment}\ }\textbf {\bibinfo
  {volume} {2018}},\ \bibinfo {pages} {083301} (\bibinfo {year}
  {2018}{\natexlab{b}})}\BibitemShut {NoStop}%
\bibitem [{\citenamefont {Tizdast}\ \emph {et~al.}(2020)\citenamefont
  {Tizdast}, \citenamefont {Ebadi}, \citenamefont {Ahadpour}, \citenamefont
  {Najafi},\ and\ \citenamefont {Mohamadzadeh}}]{tizdast2020dynamical}%
  \BibitemOpen
  \bibfield  {author} {\bibinfo {author} {\bibfnamefont {S.}~\bibnamefont
  {Tizdast}}, \bibinfo {author} {\bibfnamefont {Z.}~\bibnamefont {Ebadi}},
  \bibinfo {author} {\bibfnamefont {N.}~\bibnamefont {Ahadpour}}, \bibinfo
  {author} {\bibfnamefont {M.}~\bibnamefont {Najafi}},\ and\ \bibinfo {author}
  {\bibfnamefont {H.}~\bibnamefont {Mohamadzadeh}},\ }\href@noop {} {\bibfield
  {journal} {\bibinfo  {journal} {Physica Scripta}\ }\textbf {\bibinfo {volume}
  {95}},\ \bibinfo {pages} {115212} (\bibinfo {year} {2020})}\BibitemShut
  {NoStop}%
\bibitem [{\citenamefont {Duplantier}(2000)}]{duplantier2000conformally}%
  \BibitemOpen
  \bibfield  {author} {\bibinfo {author} {\bibfnamefont {B.}~\bibnamefont
  {Duplantier}},\ }\href@noop {} {\bibfield  {journal} {\bibinfo  {journal}
  {Physical Review Letters}\ }\textbf {\bibinfo {volume} {84}},\ \bibinfo
  {pages} {1363} (\bibinfo {year} {2000})}\BibitemShut {NoStop}%
\bibitem [{\citenamefont {Najafi}\ \emph {et~al.}(2012)\citenamefont {Najafi},
  \citenamefont {Moghimi-Araghi},\ and\ \citenamefont
  {Rouhani}}]{najafi2012avalanche}%
  \BibitemOpen
  \bibfield  {author} {\bibinfo {author} {\bibfnamefont {M.}~\bibnamefont
  {Najafi}}, \bibinfo {author} {\bibfnamefont {S.}~\bibnamefont
  {Moghimi-Araghi}},\ and\ \bibinfo {author} {\bibfnamefont {S.}~\bibnamefont
  {Rouhani}},\ }\href@noop {} {\bibfield  {journal} {\bibinfo  {journal}
  {Physical Review E}\ }\textbf {\bibinfo {volume} {85}},\ \bibinfo {pages}
  {051104} (\bibinfo {year} {2012})}\BibitemShut {NoStop}%
\bibitem [{\citenamefont {Cheraghalizadeh}\ \emph {et~al.}(2017)\citenamefont
  {Cheraghalizadeh}, \citenamefont {Najafi}, \citenamefont
  {Dashti-Naserabadi},\ and\ \citenamefont
  {Mohammadzadeh}}]{cheraghalizadeh2017mapping}%
  \BibitemOpen
  \bibfield  {author} {\bibinfo {author} {\bibfnamefont {J.}~\bibnamefont
  {Cheraghalizadeh}}, \bibinfo {author} {\bibfnamefont {M.}~\bibnamefont
  {Najafi}}, \bibinfo {author} {\bibfnamefont {H.}~\bibnamefont
  {Dashti-Naserabadi}},\ and\ \bibinfo {author} {\bibfnamefont
  {H.}~\bibnamefont {Mohammadzadeh}},\ }\href@noop {} {\bibfield  {journal}
  {\bibinfo  {journal} {Physical Review E}\ }\textbf {\bibinfo {volume} {96}},\
  \bibinfo {pages} {052127} (\bibinfo {year} {2017})}\BibitemShut {NoStop}%
\bibitem [{\citenamefont {Najafi}\ \emph {et~al.}(2021)\citenamefont {Najafi},
  \citenamefont {Tizdast},\ and\ \citenamefont
  {Cheraghalizadeh}}]{najafi2021some}%
  \BibitemOpen
  \bibfield  {author} {\bibinfo {author} {\bibfnamefont {M.}~\bibnamefont
  {Najafi}}, \bibinfo {author} {\bibfnamefont {S.}~\bibnamefont {Tizdast}},\
  and\ \bibinfo {author} {\bibfnamefont {J.}~\bibnamefont {Cheraghalizadeh}},\
  }\href@noop {} {\bibfield  {journal} {\bibinfo  {journal} {Physica Scripta}\
  }\textbf {\bibinfo {volume} {96}},\ \bibinfo {pages} {112001} (\bibinfo
  {year} {2021})}\BibitemShut {NoStop}%
\bibitem [{\citenamefont {Hentschel}\ and\ \citenamefont
  {Procaccia}(1983)}]{hentschel1983infinite}%
  \BibitemOpen
  \bibfield  {author} {\bibinfo {author} {\bibfnamefont {H.~G.~E.}\
  \bibnamefont {Hentschel}}\ and\ \bibinfo {author} {\bibfnamefont
  {I.}~\bibnamefont {Procaccia}},\ }\href@noop {} {\bibfield  {journal}
  {\bibinfo  {journal} {Physica D: Nonlinear Phenomena}\ }\textbf {\bibinfo
  {volume} {8}},\ \bibinfo {pages} {435} (\bibinfo {year} {1983})}\BibitemShut
  {NoStop}%
\bibitem [{\citenamefont {Muller}(1996)}]{muller1996characterization}%
  \BibitemOpen
  \bibfield  {author} {\bibinfo {author} {\bibfnamefont {J.}~\bibnamefont
  {Muller}},\ }\href@noop {} {\bibfield  {journal} {\bibinfo  {journal}
  {Journal of hydrology}\ }\textbf {\bibinfo {volume} {187}},\ \bibinfo {pages}
  {215} (\bibinfo {year} {1996})}\BibitemShut {NoStop}%
\bibitem [{\citenamefont {Muller}(1994)}]{muller1994characterization}%
  \BibitemOpen
  \bibfield  {author} {\bibinfo {author} {\bibfnamefont {J.}~\bibnamefont
  {Muller}},\ }\href@noop {} {\bibfield  {journal} {\bibinfo  {journal}
  {Journal of Geophysical Research: Solid Earth}\ }\textbf {\bibinfo {volume}
  {99}},\ \bibinfo {pages} {7275} (\bibinfo {year} {1994})}\BibitemShut
  {NoStop}%
\bibitem [{\citenamefont {Schreiber}\ and\ \citenamefont
  {Grussbach}(1992)}]{schreiber1992multifractal}%
  \BibitemOpen
  \bibfield  {author} {\bibinfo {author} {\bibfnamefont {M.}~\bibnamefont
  {Schreiber}}\ and\ \bibinfo {author} {\bibfnamefont {H.}~\bibnamefont
  {Grussbach}},\ }\href@noop {} {\bibfield  {journal} {\bibinfo  {journal}
  {Modern Physics Letters B}\ }\textbf {\bibinfo {volume} {6}},\ \bibinfo
  {pages} {851} (\bibinfo {year} {1992})}\BibitemShut {NoStop}%
\bibitem [{\citenamefont {Thakur}\ \emph {et~al.}(1992)\citenamefont {Thakur},
  \citenamefont {Basu}, \citenamefont {Mookerjee},\ and\ \citenamefont
  {Sen}}]{thakur1992multifractal}%
  \BibitemOpen
  \bibfield  {author} {\bibinfo {author} {\bibfnamefont {P.}~\bibnamefont
  {Thakur}}, \bibinfo {author} {\bibfnamefont {C.}~\bibnamefont {Basu}},
  \bibinfo {author} {\bibfnamefont {A.}~\bibnamefont {Mookerjee}},\ and\
  \bibinfo {author} {\bibfnamefont {A.}~\bibnamefont {Sen}},\ }\href@noop {}
  {\bibfield  {journal} {\bibinfo  {journal} {Journal of Physics: Condensed
  Matter}\ }\textbf {\bibinfo {volume} {4}},\ \bibinfo {pages} {6095} (\bibinfo
  {year} {1992})}\BibitemShut {NoStop}%
\bibitem [{\citenamefont {Chen}\ \emph {et~al.}(2019)\citenamefont {Chen},
  \citenamefont {Zhao}, \citenamefont {Jiang}, \citenamefont {Zhou},
  \citenamefont {Tian}, \citenamefont {Zeng},\ and\ \citenamefont
  {Wang}}]{chen2019detecting}%
  \BibitemOpen
  \bibfield  {author} {\bibinfo {author} {\bibfnamefont {Q.}~\bibnamefont
  {Chen}}, \bibinfo {author} {\bibfnamefont {Z.}~\bibnamefont {Zhao}}, \bibinfo
  {author} {\bibfnamefont {Q.}~\bibnamefont {Jiang}}, \bibinfo {author}
  {\bibfnamefont {J.-X.}\ \bibnamefont {Zhou}}, \bibinfo {author}
  {\bibfnamefont {Y.}~\bibnamefont {Tian}}, \bibinfo {author} {\bibfnamefont
  {S.}~\bibnamefont {Zeng}},\ and\ \bibinfo {author} {\bibfnamefont
  {J.}~\bibnamefont {Wang}},\ }\href@noop {} {\bibfield  {journal} {\bibinfo
  {journal} {Ore Geology Reviews}\ }\textbf {\bibinfo {volume} {115}},\
  \bibinfo {pages} {103182} (\bibinfo {year} {2019})}\BibitemShut {NoStop}%
\bibitem [{\citenamefont {Chhabra}\ and\ \citenamefont
  {Jensen}(1989)}]{chhabra1989direct}%
  \BibitemOpen
  \bibfield  {author} {\bibinfo {author} {\bibfnamefont {A.}~\bibnamefont
  {Chhabra}}\ and\ \bibinfo {author} {\bibfnamefont {R.~V.}\ \bibnamefont
  {Jensen}},\ }\href@noop {} {\bibfield  {journal} {\bibinfo  {journal}
  {Physical Review Letters}\ }\textbf {\bibinfo {volume} {62}},\ \bibinfo
  {pages} {1327} (\bibinfo {year} {1989})}\BibitemShut {NoStop}%
\bibitem [{\citenamefont {Jensen}\ \emph {et~al.}(1991)\citenamefont {Jensen},
  \citenamefont {Paladin},\ and\ \citenamefont
  {Vulpiani}}]{jensen1991multiscaling}%
  \BibitemOpen
  \bibfield  {author} {\bibinfo {author} {\bibfnamefont {M.~H.}\ \bibnamefont
  {Jensen}}, \bibinfo {author} {\bibfnamefont {G.}~\bibnamefont {Paladin}},\
  and\ \bibinfo {author} {\bibfnamefont {A.}~\bibnamefont {Vulpiani}},\
  }\href@noop {} {\bibfield  {journal} {\bibinfo  {journal} {Physical review
  letters}\ }\textbf {\bibinfo {volume} {67}},\ \bibinfo {pages} {208}
  (\bibinfo {year} {1991})}\BibitemShut {NoStop}%
\bibitem [{\citenamefont {Martsepp}\ \emph {et~al.}(2022)\citenamefont
  {Martsepp}, \citenamefont {Laas}, \citenamefont {Laas}, \citenamefont
  {Priimets}, \citenamefont {T{\~o}kke},\ and\ \citenamefont
  {Mikli}}]{martsepp2022dependence}%
  \BibitemOpen
  \bibfield  {author} {\bibinfo {author} {\bibfnamefont {M.}~\bibnamefont
  {Martsepp}}, \bibinfo {author} {\bibfnamefont {T.}~\bibnamefont {Laas}},
  \bibinfo {author} {\bibfnamefont {K.}~\bibnamefont {Laas}}, \bibinfo {author}
  {\bibfnamefont {J.}~\bibnamefont {Priimets}}, \bibinfo {author}
  {\bibfnamefont {S.}~\bibnamefont {T{\~o}kke}},\ and\ \bibinfo {author}
  {\bibfnamefont {V.}~\bibnamefont {Mikli}},\ }\href@noop {} {\bibfield
  {journal} {\bibinfo  {journal} {Chaos, Solitons \& Fractals}\ }\textbf
  {\bibinfo {volume} {156}},\ \bibinfo {pages} {111811} (\bibinfo {year}
  {2022})}\BibitemShut {NoStop}%
\bibitem [{\citenamefont {Tsallis}(2009)}]{tsallis2009introduction}%
  \BibitemOpen
  \bibfield  {author} {\bibinfo {author} {\bibfnamefont {C.}~\bibnamefont
  {Tsallis}},\ }\href@noop {} {\emph {\bibinfo {title} {Introduction to
  nonextensive statistical mechanics: approaching a complex world}}},\
  Vol.~\bibinfo {volume} {1}\ (\bibinfo  {publisher} {Springer},\ \bibinfo
  {year} {2009})\BibitemShut {NoStop}%
\end{thebibliography}%

\appendix
\section{Infinite Series of Fractal Dimension}~\label{SEC:Appendix}

Continuous phase transition systems possess not only local features but also global or geometric features that have been the subject of numerous theoretical studies (both analytical and simulation-based). Global observables can be employed to characterize systems at their continuous transition point, revealing hidden aspects of the models that may not be discernible through studies solely focused on local observables.  We consider here the mass pattern and define an infinite series of fractal dimensions. This section discusses the use of multifractal analysis (MA) \cite{hentschel1983infinite} for a system that contains partially filled (black) pixels. While a single set of exponents is sufficient to describe single fractal systems, this approach is not suitable for multifractal systems. In the MA approach, the space is divided into boxes of size $\delta$ \cite{muller1996characterization}. This technique has been applied in a wide range of applications, from small-scale phenomena such as chalk patterns \cite{muller1994characterization}, electronic states in Anderson localized systems \cite{schreiber1992multifractal}, and inhomogeneous potentials \cite{thakur1992multifractal}, to large-scale problems such as mountain formation \cite{chen2019detecting} and galaxy formation \cite{chhabra1989direct}. The properties of MA and its multiscaling nature \cite{jensen1991multiscaling} make it a powerful tool for analyzing complex systems.

\subsection{Mass Pattern and Box Counting}
In this analysis, we divide the system into boxes of linear size $\delta$ and examine the distribution of the filling fraction of these boxes. A pixel (or site in the model) is considered black (or occupied) if the density of mass configuration of the dried pattern at that site, denoted by $\rho_i$, is higher than the spatial average density, denoted by $\bar{\rho}$, computed over the entire sample. The spatial average density is given by the sum of densities of all pixels divided by the total number of pixels in the system, i.e., $\bar{\rho}\equiv N_{\text{pixels}}^{-1}\sum_i\rho_i$. The filling fraction of each box is determined by the number of black pixels (or occupied sites) inside the box. Specifically, if the number of black pixels in the $i$th box is $N_i (\delta)$, then the filling fraction is given by \cite{martsepp2022dependence}:
\begin{equation}
	\mu_i\equiv \frac{N_i(\delta)}{N_{\text{pixels}}}.
\end{equation}
It should be noted that the sum of the number of black pixels (or occupied sites) in all boxes, denoted by $\sum_{i=1}^{N_{\text{box}}}N_i=N_{\text{pixels}}$, is equal to the total number of pixels in the system, denoted by $N_{\text{pixels}}$. The total number of boxes is denoted by $N_{\text{box}}$. To calculate the local mass for each box and the total mass for the cluster, we can use the following method:
\begin{equation}
	m_i(\delta)\equiv 1-\delta^{k}_{N_i(\delta),0}\ , \ M(\delta)\equiv \sum_i m_i(\delta)
	\label{Eq:massFD}
\end{equation}
The Kronecker delta, represented by $\delta^{k}_{m,n}$, is a function that evaluates to $1$ if its two arguments are equal, and $0$ otherwise. It is frequently employed in mathematical and physical equations to denote the identity matrix, define functions, and describe the characteristics of vectors and tensors.

The box-counting method is a method utilized to ascertain the fractal dimension of a given system or object. This technique involves dividing the object or system into progressively smaller boxes of uniform size and calculating the number of boxes required to cover the object or system. The fractal dimension is then determined by analyzing the relationship between the size of the boxes and the number of boxes needed to cover the object or system. Therefore, the fractal dimension can be written as follows:
\begin{equation}
	D_f\equiv -\lim_{\delta\rightarrow 0} \frac{\log M_{\delta}}{\log \delta}.
\end{equation}
In a multifractal system, the fractal exponent varies depending on the scale of observation or the region of the system being considered. The multifractal analysis is a unified theory that provides a spectrum of exponents for multifractal systems. This theory is based on a generalized partition function that captures the $q$-th moment of the fluctuations of the spatial average density, denoted by $\rho_i$, in the system. The q-generalized partition function yields not only the fractal dimension but also an infinite series of fractal dimensions, including the information dimension and correlation dimension \cite{hentschel1983infinite}. So the partition function is given by the equation
\begin{equation}
	Z_q(\delta) = \sum_i\left[\mu_i(\delta) \right]^q. 
\end{equation}
The q-generalized partition function, denoted by $Z_q$, is defined as a function of the moment $q$. For scale-invariant systems, $Z_q$ scales with $\delta$ in a power-law form, but the exponent may not be a unique number at all scales. Therefore, we have:

\begin{equation}
	Z_q(\delta)\propto \delta^{\tau_q}, \ \text{so that}\ \tau_q=\lim_{\delta\rightarrow 0}\frac{\log Z_q(\delta)}{\log \delta}.
	\label{Eq:scaleInvariance}
\end{equation}

The generalized $q$-dimension is defined as follows:
\begin{equation}
	D_q\equiv \frac{\tau_q}{q-1},
\end{equation}
so that $D_f=\lim_{q\rightarrow 0}D_q$. 

\subsection{Information Dimension}
It is worth noting that if $\mu_i$ is regarded as the probability linked to a small segment ($\delta$) of the system, then the generalized $q$-dimension $D_q$ can be viewed as a normalized $q$-Renyi entropy $\mathcal{R}e_q(\delta)$ in the thermodynamic limit, as $\delta$ approaches zero, defined by
\begin{equation}
	\mathcal{R}e_q(\delta)\equiv \frac{1}{1-q}\log\sum_i\left[\mu_i(\delta) \right]^q,
\end{equation}
then
\begin{equation}
	D_q=-\lim_{\delta\rightarrow 0}	\frac{\mathcal{R}e_q(\delta)}{\log\delta}.
\end{equation}
Hence, the mass fractal dimension of samples can be linked to the Renyi entropy with q equal to zero, i.e.
\begin{equation}
	\left. \mathcal{R}e_{q=0}(\delta)\right|_{\delta\rightarrow 0}=-D_f\log \delta.
	\label{Eq:RenyLog}
\end{equation}
It is important to highlight that the scale-invariance hypothesis presented in Eq.~\ref{Eq:scaleInvariance} suggests that the Renyi entropy is proportional to the logarithm of $\delta$. Conversely, for extensive non-scale-invariant (NSI) systems, the relationship is as follows:
\begin{equation}
	\begin{split}
		& Z^{\text{NSI}}_q(\delta)=\exp\left[-f_qA\right]\\
		\rightarrow \mathcal{R}e_q^{\text{NSI}}(\delta) & =\frac{1}{q-1}\left(f_qN_{\text{boxes}}\delta^d+\text{const}\right),
	\end{split}
\end{equation}

Here, $A=N_{\text{boxes}} \delta^d$ represents the total volume of the system, which is given by the product of the number of boxes $N$-boxes and the volume of each box $\delta^d$, where $d$ is equal to $2$ in this case. It is noteworthy to compare the volume term $\delta^d$ with the logarithmic term $\log \delta$ in Eq.~\ref{Eq:RenyLog}. This logarithmic term is a key feature of scale-invariant systems, in which the system is not extensive~\cite{tsallis2009introduction}. The information dimension, which is associated with the Shannon entropy, can be defined using the following equation (note that $Z_{q=1}(\delta)=1$):
\begin{equation}
	D_1\equiv \lim_{\delta\rightarrow 0}\frac{\sum_i \mu_i(\delta)\log \mu_i(\delta)}{\log \delta}=\lim_{q\rightarrow 1}D_q
\end{equation}
In terms of the Shannon entropy 
\begin{equation}
	\mathcal{SH}(\delta)\equiv -\sum_i\mu_i\log \mu_i,
\end{equation}
we have
\begin{equation}
	\left. \mathcal{SH}(\delta)\right|_{\delta\rightarrow 0}=-D_1\log\delta.
\end{equation}

\subsection{Correlation Dimension}
Finally the correlation dimension is defined as
\begin{equation}
	\mathcal{C}\equiv \lim_{\delta\rightarrow 0}\frac{\log C(\delta)}{\log \delta},
\end{equation}
where 
\begin{equation}
	C(\delta)\equiv\frac{1}{N_{\text{pixels}}^2}\sum_{k\ne k'}\Theta(\delta-\left| \textbf{R}_k-\textbf{R}_{k'}\right| ),
\end{equation}
where $\textbf{R}_k$ is the position af the $k$th black pixel (not box), and $\Theta$ is a step function. It is shown that 
\begin{equation}
	\mathcal{C}=D_2.
\end{equation}
To see this, we note that
\begin{equation}
	\begin{split}
		\sum_{i}N_i^2 & =\sum_i\sum_{kk'}\Theta(\delta-|\textbf{R}_k-\textbf{X}_i|)\Theta(\delta-|\textbf{R}_{k'}-\textbf{X}_i|)\\
		& = \sum_i\sum_{k\ne k'}\Theta(\delta-\left| \textbf{R}_k-\textbf{R}_{k'}\right|)\delta_{\textbf{B}(\textbf{R}_k),\textbf{X}_i}\\
		&=\sum_{k\ne k'}\Theta(\delta-\left| \textbf{R}_k-\textbf{R}_{k'}\right| )=N_{\text{pixels}}^2C(\delta)
	\end{split}
\end{equation}
In the equation provided, the summation over $i$ (or $k$ and $k'$) pertains to the boxes (or pixels), where $\textbf{X}_i$ denotes the central position of the $i$-th box, and $\textbf{B}(\textbf{R})$ represents the position of the box that $\textbf{R}$ belongs to. It should be noted that the formal dimensions, as well as the higher-order dimensions, are distinct examples of the generalized dimension $D_q$, which is calculated in the second part of this section.

\section{Constructing the Distribution Function of Mass out of the Infinite Series of Fractal Dimensions}
To underscore the significance of the infinite set of fractal dimensions $D_q$ (for arbitrary values of $q$), we now examine their connection to the mass distribution function. It is worth mentioning that analogous methods can be applied to obtain other distribution functions. Initially, we focus on the monofractal system, which exhibits scale-invariant distributions, and whose moments of $N_i(\delta)$ are given by:

\begin{equation}
	\left\langle N^q\right\rangle \equiv \frac{1}{M(\delta)}\sum_{i=1}^{M(\delta)}N_i(\delta)^q=\left( \frac{N_{\text{pixels}}}{M(\delta)}\right)^qM(\delta)^{q+1}Z_q(\delta).
\end{equation}
Using Eq.~\ref{Eq:scaleInvariance} one finds
\begin{equation}
	\left\langle N^q\right\rangle=A_q\bar{N}^q\delta^{\zeta_q}
	\label{Eq:Nq}
\end{equation}
In the case of a monofractal system with scale-invariant distributions, the moments of $N_i(\delta)$ are given by the equation shown, where $A_q$ is defined as $A_q\equiv c_1^{q+1}c_2$, $\bar{N}\equiv \frac{N_{\text{pixels}}}{M(\delta)}$, and where $\bar{N}$ represents the average of $N_i$ and is equal to the total number of pixels $\frac{N_{\text{pixels}}}{M(\delta)}$. The quantity $\zeta_q\equiv \tau_q-(q+1)D_f$, and $c_1$ and $c_2$ are the proportionality constants for the $M(\delta)$-$\delta$ and $Z_q$-$\delta$ relations, respectively. The scaling moment relation is used to determine the probability distribution of $n_i$ (denoted by $p_n$). The probability characteristic function (denoted by $\tilde{p}_k$) is defined as the Fourier transform of $p_n$:
\begin{equation}
	\begin{split}
		\tilde{p}_k & \equiv \sum_{N=1}^{N_{\text{pixels}}} e^{\frac{2i\pi}{N_{\text{pixels}}} Nk}p_N=\left\langle e^{\frac{2i\pi}{N_{\text{pixels}}} Nk}\right\rangle \\
		&=\sum_{q=0}^{\infty}\frac{(2\pi ik)^q}{q!}\left\langle \left(\frac{N}{N_{\text{pixels}}}\right) ^q\right\rangle \\
		&=\tilde{p}_0\sum_{q=0}^{\infty}\frac{(2\pi i\xi k/N_{\text{pixels}})^q}{q!}\delta^{(q-1)D_q}
	\end{split}
\end{equation}
where $\xi\equiv \frac{c_1\bar{N}}{\delta^{D_f}}$, and $\tilde{p}_0\equiv\frac{c_1c_2}{\delta^{D_f}}$. Inserting Eq.~\ref{Eq:Nq} into the above function gives us
\begin{equation}
	\begin{split}
		p_N &=\frac{1}{N_{\text{pixels}}}\sum_{k=1}^{N_{\text{pixels}}}\tilde{p}_ke^{-\frac{2i\pi}{N_{\text{pixels}}}Nk}\\
		&=\sum_{q=0}^{\infty}\frac{(i\xi)^q}{q!}\delta^{(q-1)D_q}I_q(N)
	\end{split}
\end{equation}
where ($x\equiv \frac{2\pi k}{N_{\text{pixels}}}$)
\begin{equation}
	\begin{split}
		I_q(N) & \equiv \frac{\tilde{p}_0}{N_{\text{pixels}}}\sum_{k=1}^{N_{\text{pixels}}}\left( \frac{2\pi k}{N_{\text{pixels}}}\right) ^qe^{-\frac{2i\pi}{N_{\text{pixels}}}Nk}\\
		&\rightarrow \tilde{p}_0\int_0^{2\pi}\frac{\text{d}x}{2\pi} x^qe^{-ixN}\\
		&= \frac{\tilde{q}_0q!}{2\pi (iN)^{q+1}}\left(1-\frac{\Gamma[q+1,2i\pi N]}{q!}\right) ,
	\end{split}
\end{equation}
where $\Gamma[s,x]\equiv \int_x^{\infty}t^{s-1}e^{-t}\text{d}t$ is an incomplete Gamma function. Using the fact that $\Gamma(q+1,x)=q!e^{-x}\sum_{k=0}^q\frac{x^k}{k!}$, one finds
\begin{equation}
	p_N =\tilde{q}_0\sum_{q=0}^{\infty}\sum_{k=1}^q\frac{(i\xi)^q\delta^{(q-1)D_q}}{k!(-2\pi)^k(iN)^{q-k+1}}.
\end{equation}
Note also that when $|z|\to \infty$, 
\begin{equation}
	\left. \Gamma(q+1,z)\right|_{|z|\to \infty}\to z^qe^{-z}\sum_{k=0}^q \frac{q!}{(q-k)!}z^{-k}
	\label{Eq:gammaExpansion}
\end{equation}

In the given expression, the variable $z$ is a complex number and is equal to $2i\pi N$ in this context. To the first order of $1/z$and by neglecting $1$ compared to $N^q/q!$, we obtain:

\begin{equation}
	\begin{split}
		\left.I_N(q)\right|_{\text{large}\ N}&\approx i\frac{\tilde{q}_0(2\pi)^{q-1}}{N}\\
		\left. p_N\right|_{\text{large}\ N}&=\frac{i\tilde{q}_0}{2\pi N}\sum_{q=0}^{\infty}\frac{(i\xi)^q\delta^{(q-1)D_q}}{q!}.
		\label{Eq:pN}
	\end{split}
\end{equation} 
The provided equation demonstrates that the behavior of the filling fraction $N_i$ hinges on the generalized dimension $D_q$. To enhance our comprehension of this equation, we will examine a diverse spectrum of systems, where it is hypothesized that $\tau_q$ can be estimated by a smooth function with a second-order nonlinearity, expressed as:
\begin{equation}
	\tau_q=-D_f+\alpha_1q+\alpha_2q^2.
	\label{Eq:gammaq}
\end{equation}
Note that in this case we have
\begin{equation}
\begin{split}
\alpha_1 &=\frac{3}{2}D_f+2D_1-\frac{1}{2}D_2\\
\alpha_2 &=-\frac{1}{2}D_f-D_1+\frac{1}{2}D_2
\end{split}
\end{equation}
When $\alpha_2$ is zero, so that $\alpha_1=D_f+D_1$, then Eq.~\ref{Eq:pN} gives us
\begin{equation}
	\begin{split}
		\left. p_N\right|_{\text{large}\ N}&=\frac{\tilde{q}_0\delta^{-D_f}}{2\pi N}\left(ie^{2i\pi \xi \delta^{\alpha_1}} \right),\\
		&=\frac{c_1c_2\delta^{-2D_f}}{2\pi N}\left(ie^{2i\pi \xi \delta^{D_f+D_1}} \right)
	\end{split}
\end{equation}

The expression inside the parentheses in the given equation represents a pure phase, which arises from the approximations made in the derivation. In addition to this phase, we observe that a power-law dependence on $\delta$ with an exponent of $2D_f$. It is evident that the higher-order expansion terms of the incomplete gamma function require the use of higher-order dimensions $D_q$ (as seen in the higher-order terms of the expansion shown in Eq.~\ref{Eq:gammaExpansion}).

\section{Multifractal Analysis}~\label{SEC:Multi}
This section focuses on the comprehensive theory of multifractal systems and the multifractal analysis (MA) for the mass configuration of the dried pattern. Typically, multifractal systems are characterized by a range of critical exponents or fractal dimensions, in contrast to mono-fractal systems. As a result, multifractal systems comprise a \textit{spectrum of exponents}, which can be defined through the following function, representing a Legendre transformation of $\tau_q$:
\begin{equation}
	f(\alpha_q)=q\alpha_q-\tau_q,
\end{equation}
where $\alpha_q\equiv \frac{\text{d}\tau_q}{\text{d}q}$, and the exponent $\tau_q$ was defined in Eq.~\ref{Eq:scaleInvariance}. The function $f(\alpha)$ contains information about the spectrum of exponents $\tau_q$. Its peak represents the most frequently occurring exponents, and its variance indicates how much the exponents are scattered around the mean value. As an important example, let us consider Eq.~\ref{Eq:gammaq}, using which we find that
\begin{equation}
	q=\frac{\alpha_q-\alpha_1}{2\alpha_2} \rightarrow \tau_q=-D_f+\frac{1}{4\alpha_2}\left(\alpha^2_q-\alpha_1^2\right).
\end{equation}
To illustrate the behavior of $f(\alpha)$, we once again consider the smooth form given in Eq.~\ref{Eq:gammaq}, which leads to the following expression:

\begin{equation}
	f(\alpha)=D_f+\frac{1}{4\alpha_2}\left(\alpha-\alpha_1\right)^2,
\end{equation}

This indicates that the function $f(\alpha)$ has a peak around $\alpha_1$ with a width of $4\alpha_2$, and the value of $f$ at the peak point $\alpha=\alpha_1$ is equal to $D_f$. Therefore, the form of $f(\alpha)$ gives us $D_f$, $D_1$ and $D_2$. Specifically, when $\alpha_2\rightarrow 0$, $D_f$ is the amount of $f$ and the peak point, and also the peak point is $D_1+D_f$.
\end{document}